\documentclass[12pt,letterpaper]{article}
\usepackage{amsfonts}
\usepackage{amssymb}
\usepackage{amsmath}
\usepackage{amsthm}
\usepackage[round,authoryear]{natbib}
\usepackage[left=2.2cm,right=2.2cm,top=2.1cm,bottom=2.1cm]{geometry}
\usepackage{booktabs}
\usepackage{dcolumn}
\newcolumntype{d}[1]{D{.}{.}{#1}} 
\usepackage{graphicx,epstopdf}
\usepackage[T1]{fontenc}
\usepackage{lmodern}
\usepackage{enumitem}
\setlist[enumerate,1]{itemsep=1pt, topsep=1pt, partopsep=0pt, parsep=1pt}
\setlist[enumerate,2]{nosep}
\setlist[itemize,1]{itemsep=1pt, topsep=1pt, partopsep=0pt, parsep=1pt}
\setlist[itemize,2]{nosep}
\usepackage{subfig}
\usepackage{bm}
\usepackage[colorlinks,linkcolor=red,citecolor=blue,pdftex,breaklinks]{hyperref}
\hypersetup{bookmarksopen=true}
\usepackage[nameinlink]{cleveref}
\usepackage{setspace}

\theoremstyle{plain}

\theoremstyle{definition}

\Crefname{assumptionx}{Assumption}{Assumptions} 
\Crefname{examplex}{Example}{Examples} 
\Crefname{remarkx}{Remark}{Remarks} 

\makeatletter
\renewcommand*{\eqref}[1]{\hyperref[{#1}]{\textup{\tagform@{\ref*{#1}}}}}
\makeatother

\makeatletter
\DeclareRobustCommand\citepos
  {\begingroup\def\NAT@nmfmt##1{{\NAT@up##1's}}%
   \NAT@swafalse\let\NAT@ctype\z@\NAT@partrue
   \@ifstar{\NAT@fulltrue\NAT@citetp}{\NAT@fullfalse\NAT@citetp}}
\makeatother

\expandafter
\def \expandafter \normalsize \expandafter{\normalsize \setlength \abovedisplayskip{8pt plus 2pt minus 7pt}}
\expandafter
\def \expandafter \normalsize \expandafter{\normalsize \setlength \abovedisplayshortskip{0pt plus 2pt}}
\expandafter
\def \expandafter \normalsize \expandafter{\normalsize \setlength \belowdisplayskip{8pt plus 2pt minus 7pt}}
\expandafter
\def \expandafter \normalsize \expandafter{\normalsize \setlength \belowdisplayshortskip{3pt plus 1pt minus 2pt}}

\def\bbeta{{\bm\beta}}
\def\bgamma{{\bm\gamma}}

\def\biX{{\bm{X}}}
\def\biZ{{\bm{Z}}}

\def\biM{{\bm{M}}}
\def\biP{{\bm{P}}}
\def\biA{{\bm{A}}}

\def\biH{{\bm{H}}}
\def\bih{{\bm{h}}}
\def\biR{{\bm{R}}}

\def\bir{{\bm{r}}}
\def\bis{{\bm{s}}}
\def\biy{{\bm{y}}}
\def\bix{{\bm{x}}}

\def\biu{{\bm{u}}}

\def\bSigma{{\bm{\Sigma}}}
\def\biV{{\bm{V}}}

\def\bdelta{{\bm{\delta}}}

\def\E{{\rm E}}

\def\IF{{\mathbb I}}

\def\tk{\kern 0.08333em}
\def\tn{\kern -0.08333em}
\def\tkk{\kern 0.04167em}
\def\bzero{{\bm{0}}}
\def\bfI{{\bf I}}

\def\th#1{$#1^{\tk{\rm th}}$}

\DeclareMathOperator{\var}{Var}

\usepackage[small,compact]{titlesec}

\begin{document}

\title{Fast and Reliable Jackknife and Bootstrap\\ Methods for
Cluster-Robust Inference\thanks{We are grateful to David Drukker,
Alexander Fischer, David Roodman, the Co-Editor, Francis Vella, an 
anonymous referee, and seminar participants at Aarhus University, 
Carleton University, University of Toronto, and New York Camp 
Econometrics 2022 for helpful comments and suggestions.
MacKinnon and Webb thank the Social Sciences and Humanities Research
Council of Canada (SSHRC grants 435-2016-0871 and 435-2021-0396) for
financial support. Nielsen thanks the Danish National Research
Foundation for financial support (DNRF Chair grant number DNRF154).}}

\author{James G. MacKinnon\thanks{Corresponding author. Address: 
Department of Economics, 94 University Avenue, Queen's University, 
Kingston, Ontario K7L 3N6, Canada. Email:\
\texttt{mackinno@queensu.ca}. Tel.\ 613-533-2293. Fax
613-533-6668.}\\Queen's University\\ \texttt{mackinno@queensu.ca} \and
Morten \O rregaard Nielsen\\Aarhus University\\
\texttt{mon@econ.au.dk} \and Matthew D. Webb\\Carleton
University\\ \texttt{matt.webb@carleton.ca}}

\maketitle

\begin{abstract}
We provide computationally attractive methods to obtain 
jackknife-based cluster-robust variance matrix estimators
(CRVEs) for linear regression models estimated by least squares. We
also propose several new variants of the wild cluster bootstrap, which
involve these CRVEs, jackknife-based bootstrap data-generating
processes, or both. Extensive simulation experiments suggest that the
new methods can provide much more reliable inferences than existing
ones in cases where the latter are not trustworthy, such as when the
number of clusters is small and/or cluster sizes vary substantially.
Three empirical examples illustrate the new methods.

\vskip 12pt



\end{abstract}

\clearpage
\onehalfspacing

\section{Introduction}
\label{sec:intro}

In applications of linear regression models to many fields of
economics and other disciplines, it is common to divide the sample
into disjoint clusters and employ a cluster-robust variance matrix
estimator (or CRVE) for inference. These estimators are based on the
assumption that the disturbances of the regression model are
uncorrelated across clusters, but they allow for arbitrary patterns of
dependence and heteroskedasticity within each cluster. The literature
on cluster-robust inference has grown rapidly in recent years.
\citet{CM_2015} is a classic survey article. \citet*{CGH_2018} surveys
a broader class of methods for dependent data. \citet*{MNW-guide}
provides a guide that explores the implications of key theoretical
results for empirical practice, with an emphasis on bootstrap methods.

There are several CRVEs for ordinary least squares (OLS) estimates of
linear regression models; see \Cref{sec:model}. However, because the
one known as CV$_{\tn1}$ is easy to compute and is the default in
\texttt{Stata}, almost all empirical work to date has used it. 
Cluster-robust tests and confidence intervals based on CV$_{\tn1}$ may 
or may not yield reliable inferences. Whether they do so depends 
primarily on the number of clusters $G$ and how homogeneous these are. 
When all clusters are roughly equal in size and approximately balanced,
asymptotic inference based on CV$_{\tn1}$ seems to be fairly reliable
whenever $G$ is at least moderately large (say 50 or more). However,
even when $G$ is very large, cluster-robust $t$-tests and Wald tests
are at risk of severe over-rejection, and cluster-robust confidence
intervals are at risk of severe under-coverage in at least two
situations. The first is when one or a few clusters are much larger
than the rest, and the second is when the only ``treated''
observations belong to just a few clusters; \citet*{DMN_2019}
discusses the first case, and \citet*{MW-JAE,MW-EJ} discuss the
second.

Alternatives to CV$_{\tn1}$ have been known since \citet{BM_2002}. The
first contribution of the present paper, which is discussed in
\Cref{sec:jack}, is to provide a fast method for computing
jackknife\tkk-based CRVEs, of which the simplest is generally known as
CV$_{\tn3}$. By explicitly using the cluster jackknife for computation
in an efficient way, our method makes it feasible to employ
CV$_{\tn3}$ for inference even in very large samples with a large
number of clusters.

Because CV$_{\tn3}$ standard errors used to be hard to compute, there
has been very little work comparing the finite\tkk-sample performance
of $t$-tests based on CV$_{\tn3}$ with those of similar procedures
based on CV$_{\tn1}$; a partial exception is \citet{NW_2022}. The
second contribution of this paper is to compare the finite\tkk-sample
properties of these tests, and also ones based on CV$_{\tn2}$, by 
simulation; see \Cref{sec:sims}. In concurrent work that 
cites our simulations, \citet{Hansen-jack} provides important 
theoretical results which suggest that asymptotic inference based on 
CV$_{\tn3}$ is generally more reliable, and more conservative, than 
asymptotic inference based on CV$_{\tn1}$.

Existing bootstrap methods for cluster-robust inference are all based
on CV$_{\tn1}$. The best known of these (and until now the best
performing one) seems to be the wild cluster restricted (or WCR)
bootstrap proposed in \citet*{CGM_2008}. There is also a closely
related procedure called the wild cluster unrestricted (or WCU)
bootstrap, which generally does not work quite as well. The asymptotic
validity of these procedures is proved in \citet{DMN_2019}, which also
analyzes their higher-order asymptotic properties. Until a few years
ago, the WCR and WCU bootstraps were computationally expensive for
large samples, but that is no longer the case. \citet*{RMNW} describes
a remarkably efficient implementation in the \texttt{Stata} package
\texttt{boottest}, and \citet{JGM-fast} discusses other methods for
fast computation. The \texttt{boottest} routines are now available as
a \texttt{Julia} package which can be also be called from \texttt{R},
\texttt{Python}, and \texttt{Stata}. The package
\texttt{fwildclusterboot} implements the \texttt{boottest} method
natively in \texttt{R} \citep{Fischer_2022}.

The third contribution of this paper is to propose several new
variants of the wild cluster bootstrap. One modification simply
replaces CV$_{\tn1}$ by CV$_{\tn3}$. The other, which requires some
new results, involves modifying the bootstrap data-generating process,
or DGP. Modern treatments of the wild cluster bootstrap, such as
\citet{MNW-guide}, express the bootstrap DGP as a function of the
empirical scores. We show how to make the bootstrap DGP more closely
resemble the (unknown) true DGP by transforming the residuals before
forming the scores. The transformation we propose is based on the
jackknife. Accordingly, it does not actually require any calculations
that explicitly involve residuals. This makes it very fast when the
number of clusters is small relative to the sample size, even when the
latter is extremely large.

The next section establishes notation and briefly reviews the
literature on asymptotic cluster-robust inference for the linear
regression model. \Cref{sec:jack} then discusses a computational
method for CV$_{\tn3}$ which is conceptually simple and extremely
fast in many cases, as we demonstrate in \Cref{sec:speed}. Next,
\Cref{sec:boot} discusses several ways of modifying the wild cluster
bootstrap. Simulation results in \Cref{sec:sims} suggest that our new
versions of the WCR and WCU bootstraps perform better, sometimes very
much better, than the original ones. This is particularly true when
cluster sizes vary greatly. One modified version of the WCR bootstrap
that uses transformed scores seems to work especially well in most
settings. \Cref{sec:empirical} presents three empirical examples in
which our methods are likely to be more reliable than existing ones. 
\Cref{sec:conc} concludes with a brief discussion of the methods that 
we recommend in practice.

\section{The Linear Regression Model with Clustering}
\label{sec:model}

Consider the linear regression model $y_i = \bix_i^\top\!\bbeta +
u_i$. If we divide the data into $G$ disjoint clusters, where the
allocation of observations to clusters is assumed to be known, this
can be written as
\begin{equation}
\label{eq:lrmodel} 
\biy_g =\biX_g\bbeta + \biu_g, \quad g=1,\ldots,G.
\end{equation}
The \th{g} cluster has $N_g$ observations, and the total sample size is
$N = \sum_{g=1}^G N_g$. In \eqref{eq:lrmodel}, $\biX_g$ is an
$N_g\times k$ matrix of regressors, $\bbeta$ is a $k$-vector of
coefficients, $\biy_g$ is an $N_g$-vector of observations on the
regressand, and $\biu_g$ is an $N_g$-vector of disturbances (or error
terms). Stacking the $\biy_g$ yields the $N$-vector $\biy$, stacking
the $\biX_g$ yields the $N\times k$ matrix $\biX$\tn, and stacking the
$\biu_g$ yields the $N$-vector $\biu$, so that \eqref{eq:lrmodel} can
be rewritten as $\biy = \biX\tn\bbeta + \biu$.

The OLS estimator of $\bbeta$ is
\begin{equation}
\hat\bbeta = (\biX^\top\!\biX)^{-1}\biX^\top\biy
= \bbeta_0 +  (\biX^\top\!\biX)^{-1}\biX^\top\biu,
\label{eq:OLSbeta}
\end{equation}
where the second equality depends on the assumption that the data are
actually generated by \eqref{eq:lrmodel} with true value $\bbeta_0$.
Thus, if $\bis_g = \biX_g^\top \biu_g$ is the score vector for the
\th{g} cluster,
\begin{equation}
\hat\bbeta - \bbeta_0 = (\biX^\top\!\biX)^{-1}\sum_{g=1}^G \biX_g^\top\biu_g
= \Big(\tn\sum_{g=1}^G\biX_g^\top\!\biX_g\Big)^{\!\!-1} \sum_{g=1}^G \bis_g.
\label{eq:betahat}
\end{equation}
Obtaining valid inferences evidently requires assumptions about the
score vectors. For a correctly specified model, $\E(\bis_g)=\bzero$
for all $g$. We further assume that
\begin{equation}
\label{eq:Sigma_g}
\E(\bis_g\bis_g^\top) = \bSigma_g \quad\mbox{and}\quad
\E(\bis_g\bis_{g'}^\top) = \bzero, \quad g,g'=1,\ldots,G,\quad g'\ne g,
\end{equation}
where $\bSigma_g$ is the symmetric, positive semidefinite variance
matrix of the scores for the \th{g} cluster. The second assumption in
\eqref{eq:Sigma_g} is crucial. It states that the scores for every
cluster are uncorrelated with the scores for every other cluster.

From the rightmost expression in \eqref{eq:betahat}, we see that the
distribution of $\hat\bbeta$ depends on the disturbance subvectors
$\biu_g$ only through the distribution of the score vectors $\bis_g$.
It follows immediately that an estimator of $\var(\hat\bbeta)$ should
be based on the usual sandwich formula,
\begin{equation}
\label{eq:trueV}
(\biX^\top\!\biX)^{-1} \Big(\tn\sum_{g=1}^G \bSigma_g\tn\Big) 
(\biX^\top\!\biX)^{-1}.
\end{equation}
Every CRVE replaces the $\bSigma_g$ in \eqref{eq:trueV} by functions
of the $\biX_g$ and the residual subvectors $\hat\biu_g$. There is
more than one way to do this. Since $\bSigma_g$ is the expectation of
$\bis_g\bis_g^\top$, the simplest approach is just to replace it by
$\hat\bis_g\hat\bis_g^\top$, where $\hat\bis_g = \biX_g^\top
\hat\biu_g$ is the empirical score vector for the \th{g} cluster. If
in addition we multiply by a correction for degrees of freedom, we
obtain
\begin{equation}
\mbox{CV$_{\tn1}$:}\qquad \hat\biV_1(\hat\bbeta) =
\frac{G(N-1)}{(G-1)(N-k)}
(\biX^\top\!\biX)^{-1}
\Big(\tn\sum_{g=1}^G \hat\bis_g\hat\bis_g^\top\Big) (\biX^\top\!\biX)^{-1}.
\label{eq:CV1}
\end{equation}
This is by far the most widely-used CRVE in practice, and it is the
default in \texttt{Stata}. The leading scalar is chosen so that, when
$G=N$\tn, $\hat\biV_1(\hat\bbeta)$ reduces to the familiar HC$_1$
estimator \citep{MW_1985} that is robust only to heteroskedasticity of
unknown form.

Inference typically relies on cluster-robust $t$-statistics and Wald
statistics based on \eqref{eq:CV1}. If $\beta_j$ denotes the \th{j}
element of $\bbeta$, $\hat\beta_j$ the OLS estimate, and $\beta_{0j}$
its value under the null hypothesis, then the appropriate
$t$-statistic is
\begin{equation}
t_j = \frac{\hat\beta_j - \beta_{0j}}{\textrm{se}_1(\hat\beta_j)},
\label{eq:cr-tstat}
\end{equation}
where $\textrm{se}_1(\hat\beta_j)$ is the square root of the \th{j}
diagonal element of $\hat\biV_1(\hat\bbeta)$. Under extremely strong
assumptions, \citet*{BCH_2011} shows that $t_j$ asymptotically follows
the $t(G-1)$ distribution. Conventional ``asymptotic'' inference is
based on this distribution.

We should expect inferences based on CV$_{\tn1}$ to be reliable if the
sum of the $\bis_g$, suitably normalized, is well approximated by a
multivariate normal distribution with mean zero, and if the $\bis_g$
are well approximated by the $\hat\bis_g$. But asymptotic inference
can be misleading when either or both of these approximations is poor;
see \citet{DMN_2019} and \citet{MNW-guide}. Whether or not the first 
approximation is a good one depends on the model and the data, and 
there is not much the investigator can do about it. But the second 
approximation can, in principle, be improved by using modified empirical
score vectors instead of the $\hat\bis_g$.

Two CRVEs based on this idea, usually known as CV$_{\tn2}$ and
CV$_{\tn3}$, were proposed (under different names) in \citet{BM_2002}.
These are the cluster analogs of the hetero\-skedasticity-consistent
variance matrix estimators HC$_2$ and HC$_3$ proposed in
\citet{MW_1985}. All of these estimators are designed to compensate,
in different ways, for the shrinkage and intra-cluster correlation of
the residuals induced by least squares.

The CV$_{\tn2}$ variance matrix is
\begin{equation}
\mbox{CV$_{\tn2}$:}\qquad \hat\biV_2(\hat\bbeta) =
(\biX^\top\biX)^{-1}\Big(\tk\sum_{g=1}^G \grave\bis_g\grave\bis_g^\top
\Big)(\biX^\top\biX)^{-1},
\label{eq:CV2}
\end{equation}
where the modified score vectors $\grave\bis_g$ are defined as
\begin{equation}
\grave\bis_g = \biX_g^\top\biM_{gg}^{-1/2}\tk\hat\biu_g.
\label{eq:bis2}
\end{equation}
Here $\biM_{gg} = \bfI_{N_g} -
\biX_g(\biX^\top\biX)^{-1}\tn\biX_g^\top$ is the \th{g} diagonal block
of the projection matrix $\biM_\biX$, which satisfies $\hat\biu =
\biM_\biX\biu$, and $\biM_{gg}^{-1/2}$ is the symmetric square root of
its inverse. The CV$_{\tn2}$ estimator has been recommended in
\citet{Imbens_2016} and \citet{PT_2018}. Following \citet{BM_2002},
these papers provide methods for computing critical values based on
$t$ and $F$ distributions with computed degrees of freedom.

The CV$_{\tn3}$ variance matrix is very similar to CV$_{\tn2}$, but,
as we explain in \Cref{sec:jack}, it is based on the jackknife. The
usual definition is
\begin{equation}
\mbox{CV$_{\tn3}$:}\qquad \hat\biV_3(\hat\bbeta) = \frac{G-1}{G}
(\biX^\top\biX)^{-1}\Big(\tk\sum_{g=1}^G \acute\bis_g\acute\bis_g^\top
\Big)(\biX^\top\biX)^{-1},
\label{eq:CV3}
\end{equation}
where now the modified score vectors $\acute\bis_g$ are defined as
\begin{equation}
\acute\bis_g = \biX_g^\top\biM_{gg}^{-1}\tk\hat\biu_g.
\label{eq:bis3}
\end{equation}
The rescaling factor $(G-1)/G$ in \eqref{eq:CV3} is the analog of the
factor $(N-1)/N$ that occurs in jackknife variance matrix estimators
at the individual level. This factor implicitly assumes that all
clusters are the same size and perfectly balanced, with disturbances
that are independent and homoskedastic; an alternative is proposed in
\citet{NW_2022}.

Computing either CV$_{\tn2}$ or CV$_{\tn3}$ using \eqref{eq:CV2} or
\eqref{eq:CV3} is extremely expensive, or even computationally
infeasible, when any of the $N_g$ are large. The problem is that,
before computing \eqref{eq:bis3}, we apparently need to rescale the
residual vector $\hat\biu_g$ for each cluster. This involves storing
and inverting the $N_g\times N_g$ matrix $\biM_{gg}$. Before computing
\eqref{eq:bis2}, we also need to compute the symmetric square roots of
the $\biM_{gg}$, and this requires calculating their eigenvalues and
eigenvectors. Of course, when all clusters are very small, this is not
difficult. When $G=N$\tn, CV$_{\tn2}$ reduces to HC$_2$, and
CV$_{\tn3}$ reduces to HC$_3$, both of which can be computed very
quickly.

\citet{NAAMW_2020} has recently proposed a method that is much faster
for large clusters. Versions of this method apply to both CV$_{\tn2}$
and CV$_{\tn3}$. Instead of rescaling the residual vectors, it
calculates the score vectors $\grave\bis_g$ or $\acute\bis_g$ directly
using equations that do not involve any $N_g\times N_g$ matrices. A
revised version of this method, which appears to be new, works as 
follows. First, form the $k\times k$ matrices
\begin{equation}
\biA_g = (\biX^\top\biX)^{-1/2}\biX_g^\top\biX_g(\biX^\top\biX)^{-1/2},
\quad g=1,\ldots,G.
\label{def:A}
\end{equation}
Then, for \eqref{eq:CV2}, calculate the rescaled score vectors
\begin{equation}
\grave\bis_g = (\biX^\top\biX)^{1/2}(\bfI_k - \biA_g)^{-1/2}
(\biX^\top\biX)^{-1/2}\hat\bis_g, \quad g=1,\ldots,G,
\label{eq:bis2f}
\end{equation}
and, for \eqref{eq:CV3}, calculate the rescaled score vectors
\begin{equation}
\acute\bis_g = (\biX^\top\biX)^{1/2}(\bfI_k - \biA_g)^{-1}
(\biX^\top\biX)^{-1/2}\hat\bis_g, \quad g=1,\ldots,G.
\label{eq:bis3f}
\end{equation}
These rescaled score vectors are used in \eqref{eq:CV2} and
\eqref{eq:CV3} as before. Unless all the clusters are very small,
computing CV$_{\tn2}$ and CV$_{\tn3}$ using \eqref{eq:bis2f} and
\eqref{eq:bis3f} is much faster than computing them using
\eqref{eq:bis2} and \eqref{eq:bis3}. In the case of CV$_{\tn3}$, an
even faster and more intuitive method is available. This
jackknife\tkk-based method, which we discuss in the next section, can
be extremely fast when $N$ is large and $G$ is much smaller than
$N$\tn, so that at least some clusters are large; see
\Cref{sec:speed}.

\section{Jackknife Variance Matrix Estimators}
\label{sec:jack}

The jackknife is a simple method for reducing bias and estimating
standard errors by omitting observations sequentially.
\citet{Tukey_1958} suggested using the jackknife to estimate standard
errors, and \citet{Miller_1974} is a classic reference.  The key idea
of the cluster jackknife is to compute $G$ sets of parameter
estimates, each of which omits one cluster at a time. In this section,
we use it to compute two closely related CRVEs.

The OLS estimates of $\bbeta$ when each cluster is omitted in turn are
\begin{equation}
\label{eq:delone}
\hat\bbeta^{(g)} = (\biX^\top\!\biX - \biX_g^\top\!\biX_g)^{-1}
(\biX^\top\biy - \biX_g^\top\biy_g), \quad g=1,\ldots,G.
\end{equation}
It is easy to obtain the $\hat\bbeta^{(g)}$ in a computationally
efficient manner. We start by calculating the cluster-level matrices
and vectors
\begin{equation}
\biX_g^\top\!\biX_g \quad\mbox{and}\quad \biX_g^\top\biy_g, \quad
g=1,\ldots,G.
\label{eq:subthings}
\end{equation}
Unless $G$ is very large, this involves very little cost beyond that
of computing $\hat\bbeta$, because we can use the quantities in
\eqref{eq:subthings} to construct $\biX^\top\!\biX$ and
$\biX^\top\biy$ and then use \eqref{eq:OLSbeta} to obtain
$\hat\bbeta$. For typical values of $k$, it should then be reasonably
inexpensive to calculate $\hat\bbeta^{(g)}$ for every cluster using
\eqref{eq:delone}. The main cost, beyond that of computing
$\hat\bbeta$, is that we need to calculate the inverse of a $k\times
k$ matrix for each of the $\hat\bbeta^{(g)}$\tn.

The cluster jackknife estimator of $\var(\hat\bbeta)$ is the cluster
analog of the usual jackknife variance matrix estimator given in
\citet{Efron_81}, among others. It is defined as
\begin{equation}
\mbox{CV$_{\tn3{\rm J}}$:}\qquad \hat\biV_{3{\rm J}}(\hat\bbeta) =
\frac{G-1}{G} \sum_{g=1}^G (\hat\bbeta^{(g)} -
\bar\bbeta)(\hat\bbeta^{(g)} - \bar\bbeta)^\top,
\label{eq:jack}
\end{equation}
where $\bar\bbeta = G^{-1}\sum_{g=1}^G \hat\bbeta^{(g)}$ is the sample
average of the $\hat\bbeta^{(g)}$. Notice that \eqref{eq:jack}
calculates the variance matrix around $\bar\bbeta$. Centering around
$\bar\bbeta$ is common in jackknife variance estimation, but it is
also common to center around $\hat\bbeta$, as in \citet{BM_2002}.

There is a very close relationship between $\hat\biV_{3{\rm J}}
(\hat\bbeta)$ and $\hat\biV_3(\hat\bbeta)$. In fact,
\begin{equation}
\hat\biV_3(\hat\bbeta) = \frac{G-1}{G} \sum_{g=1}^G
(\hat\bbeta^{(g)} - \hat\bbeta)(\hat\bbeta^{(g)} - \hat\bbeta)^\top,
\label{eq:jack3}
\end{equation}
which is just \eqref{eq:jack} with $\bar\bbeta$ replaced by
$\hat\bbeta$. This follows from \eqref{eq:CV3} and \eqref{eq:bis3}
because
\begin{equation}
(\biX^\top\biX)^{-1}\acute\bis_g =
(\biX^\top\biX)^{-1}\biX_g^\top\biM_{gg}^{-1}\hat\biu_g = \hat\bbeta
- \hat\bbeta^{(g)}.
\label{eq:keyres}
\end{equation}
Note that the summation in \eqref{eq:jack3} is unchanged if
$\hat\bbeta^{(g)} - \hat\bbeta$ is replaced by $\hat\bbeta -
\hat\bbeta^{(g)}$.

Although the second equality in \eqref{eq:keyres} is not new, it will
turn out to be very useful in \Cref{sec:boot}, and so we now prove it.
The middle expression in \eqref{eq:keyres} can be written as
\begin{equation}
(\biX^\top\biX)^{-1}\biX_g^\top\biM_{gg}^{-1}\biy_g -
(\biX^\top\biX)^{-1}\biX_g^\top\biM_{gg}^{-1}\biX_g
(\biX^\top\biX)^{-1}\biX^\top\biy.
\label{eq:int1}
\end{equation}
Using the Woodbury matrix identity,
\begin{equation}
(\biX^\top\biX - \biX_g^\top\biX_g)^{-1}
= (\biX^\top\biX)^{-1} +
(\biX^\top\biX)^{-1}\biX_g^\top\biM_{gg}^{-1}\biX_g(\biX^\top\biX)^{-1},
\label{eq:update}
\end{equation}
$\hat\bbeta^{(g)}$ can be written as the sum of four terms, the first
of which is just $\hat\bbeta$. Thus the right-hand side of
\eqref{eq:keyres} can be written as
\begin{equation}
\begin{aligned}
&(\biX^\top\biX)^{-1}\biX_g^\top\biM_{gg}^{-1}\biX_g
(\biX^\top\biX)^{-1}\biX_g^\top\biy_g
+ (\biX^\top\biX)^{-1}\biX_g^\top\biy_g\\
&\qquad-(\biX^\top\biX)^{-1}\biX_g^\top\biM_{gg}^{-1}\biX_g
(\biX^\top\biX)^{-1}\biX^\top\biy.
\label{eq:int2}
\end{aligned}
\end{equation}
The last term in \eqref{eq:int2} is identical to the last term in 
\eqref{eq:int1}. The first two terms in \eqref{eq:int2} can be
rewritten as
\begin{equation*}
(\biX^\top\biX)^{-1}\biX_g^\top\biM_{gg}^{-1}\biP_{gg}\biy_g
+ (\biX^\top\biX)^{-1}\biX_g^\top\biy_g,
\end{equation*}
where $\biP_{gg} = \biX_g (\biX^\top\biX)^{-1}\biX_g^\top$ is the 
\th{g} diagonal block of the matrix $\biP_\biX = \bfI - \biM_\biX$, so
that $\biP_{gg} = \bfI-\biM_{gg}$. Inserting this straightforwardly 
yields the result that
\begin{equation}
\begin{aligned}
&(\biX^\top\biX)^{-1}\biX_g^\top\biM_{gg}^{-1}\biP_{gg}\biy_g
+ (\biX^\top\biX)^{-1}\biX_g^\top\biy_g\\
&\quad= (\biX^\top\biX)^{-1}\biX_g^\top\biM_{gg}^{-1}(\bfI-\biM_{gg})\biy_g
+ (\biX^\top\biX)^{-1}\biX_g^\top\biy_g 
= (\biX^\top\biX)^{-1}\biX_g^\top\biM_{gg}^{-1}\biy_g.
\end{aligned}
\label{eq:lastline}
\end{equation}
The right-hand side of \eqref{eq:lastline} is the first term in
\eqref{eq:int1}, which proves the second equality
in~\eqref{eq:keyres}. When $N_g=1$ for all $g$, $\hat\biV_{3{\rm
J}}(\hat\bbeta)$ is numerically equal to the original HC$_3$ estimator
proposed in \citet{MW_1985}. The modern version of HC$_3$, which uses
$\hat\bbeta$ instead of $\bar\bbeta$ and omits the factor of
$N/(N-1)$, is due to \citet[Chapter~16]{DM_1993}.

Both cluster jackknife estimators may be used to compute
cluster-robust $t$-statistics. Since there are $G$ terms in the
summation, it is natural to compare these with quantiles of the
$t(G-1)$ distribution, as usual. These procedures should almost always
be more conservative than $t$-tests based on CV$_{\tn1}$
\citep{Hansen-jack}. We expect CV$_{\tn3}$ and CV$_{\tn3{\rm{J}}}$ to
be very similar in most cases. This issue will be investigated in
\Cref{subsec:size}, where we conclude that it is reasonable to focus
on CV$_{\tn3}$.

Both CV$_3$ and CV$_{\tn3{\rm J}}$ have been available in \texttt{Stata}
for some years by using the options ``\texttt{vce(jackknife,mse)}'' and 
``\texttt{vce(jackknife)}'', respectively. However, the 
implementations discussed here are much more efficient when $G$ is not 
very small. They are available in \texttt{Stata} and \texttt{R} 
packages, both named \texttt{summclust}; see \cite*{MNW_summclust} and 
\cite{Fischer_summclust}. Both packages also calculate a number of 
summary statistics that may be used to assess the reliability of 
cluster-robust inference as described in \cite*{MNW-influence}.

The matrix $\biX^\top\!\biX - \biX_g^\top\!\biX_g$ in
\eqref{eq:delone} can be singular for one or more values of~$g$, so
that at least some elements of $\hat\bbeta^{(g)}$ cannot be
identified. This can happen in otherwise well-specified models when
there are cluster-level fixed effects. In that case, the solution is
simply to partial them out before running the regression. In other
cases where a singularity occurs, there are two possible courses of
action. The first is to modify \eqref{eq:jack} and \eqref{eq:jack3} so
that the summation is taken only over values of $g$ for which
$\hat\bbeta^{(g)}$ can be estimated, and $G$ is replaced by the number
of clusters for which that is the case (this is the approach followed
in the native \texttt{Stata} implementations; see also
\Cref{subsec:patronage}). When there are only a few problematic 
clusters, this approach may be attractive. But since $\hat\bbeta$ and 
$\bar\bbeta$ would then be based on different samples, it seems likely
that CV$_{\tn3{\rm J}}$ and CV$_{\tn3}$ may differ more than they
would usually do, which suggests that it may be safer to use the
former.

The second course of action is to replace the inverse in 
\eqref{eq:delone} by a generalized inverse. In practice, this means
replacing coefficients that cannot be identified by zeros. When the
elements of $\hat\bbeta^{(g)}$ that are of primary interest can always
be identified, this approach may be attractive, especially when there
are many problematic clusters, as in the example of
\Cref{subsec:patronage}.

\section{Speed of Computation}
\label{sec:speed}

The CV$_{\tn3}$ estimator can be challenging to compute. Following
\citet{BM_2002}, it is natural to employ what we call the ``residual
method'' based on \eqref{eq:CV3} and \eqref{eq:bis3}. To compute the
modified score vector $\acute\bis_g$ for the \th{g} cluster, this
method uses the $N_g$-vector of residuals $\hat\biu_g$ and the
$N_g\times N_g$ matrix $\biM_{gg}^{-1}$. Unless every $N_g$ is small,
storing and inverting the $\biM_{gg}$ matrices is computationally
expensive. Indeed, for even moderately large values of the $N_g$, this
can be effectively impossible.

A much faster method, recently proposed in \citet{NAAMW_2020} and
revised modestly in \Cref{sec:model}, uses \eqref{eq:bis3f} to obtain
the modified score vectors $\acute\bis_g$. Since it operates directly
on the score vectors $\hat\bis_g$, we call it the ``score method.'' An
even faster approach, discussed in \Cref{sec:jack}, computes the
$\hat\bbeta^{(g)}$ using \eqref{eq:delone} and then calculates their
variance matrix as \eqref{eq:jack3}. For obvious reasons, we refer to
this as the ``jackknife method.''

To compare timings for the residual, score, and jackknife methods, we 
generate two datasets with $N = 1,048,576 = 2^{20}$ observations. In 
one case, there are 20 regressors, and in the other case there are~40.
The observations are divided into $G$ equal-sized clusters, where $G$ 
varies from 16 to $512K$ and $K$ denotes $1024=2^{10}$\tn. Thus the 
cluster size $M=N/G$ varies from 2 to~$64K$.

\begin{figure}[tb]
\begin{center}
\caption{Timings for three ways to compute CV$_{\tn 3}$}
\vspace*{-0.5em}
\label{fig:1}
\includegraphics[width=0.80\textwidth]{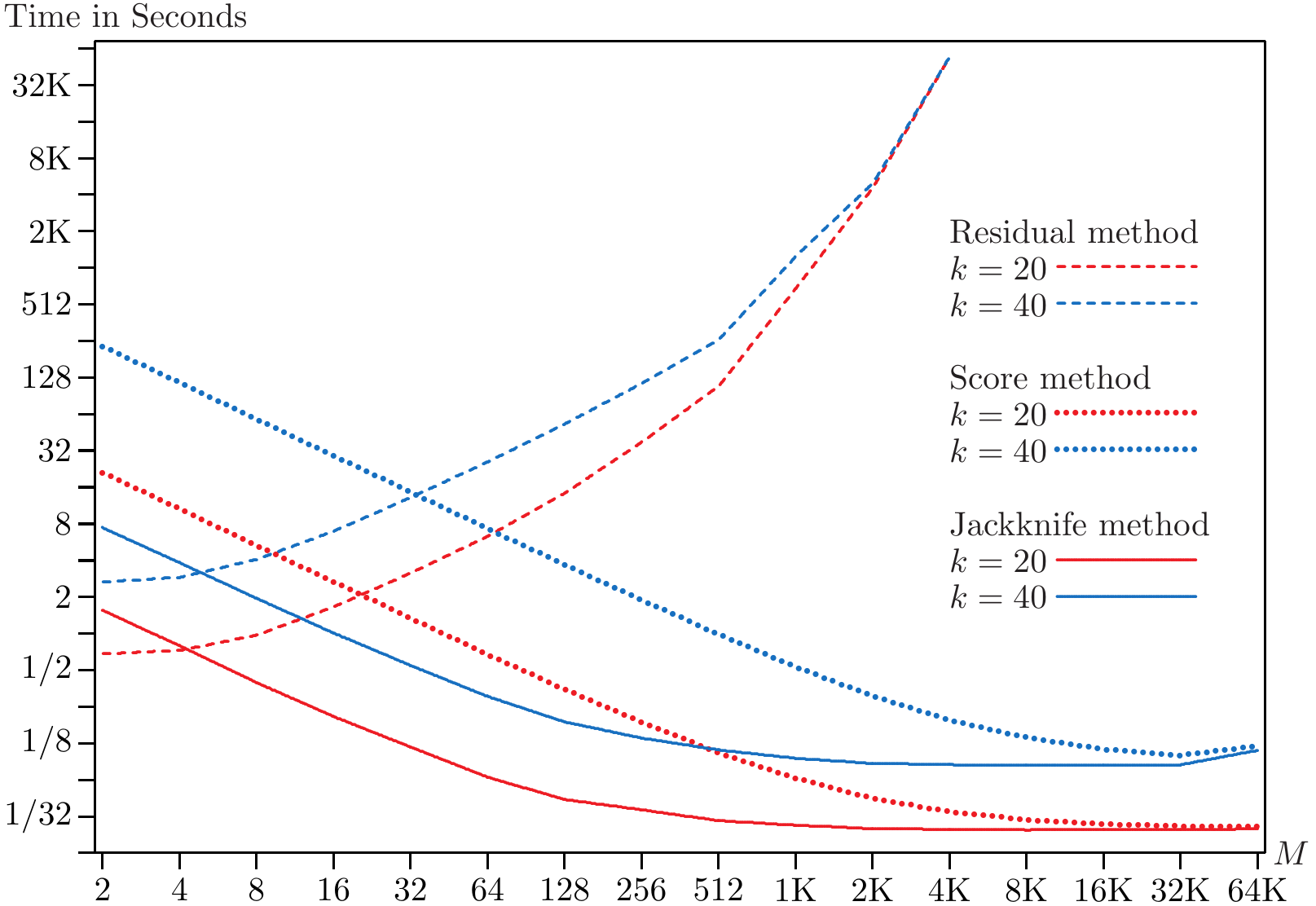}
\end{center}
{\footnotesize \textbf{Notes:} The sample size is $N=1,048,576=1024K$,
where $K=1024=2^{10}$\tn. The number of clusters varies from 16 
to~$512K$. All clusters have $M=N/G$ observations, so that cluster sizes
vary from 2 to $64K$. The number of regressors $k$ is either 20 or~40. 
Times required to compute $\hat\bbeta$ are included; see text. All
computations were performed in Fortran using one core of an Intel
i9-13900K processor.}
\end{figure}

\Cref{fig:1} shows the time in seconds, on a log$_2$ scale, for each
of the three methods and the two datasets as a function of cluster
sizes $M=N/G$, which vary from 2 to~$64K$\tn. These times include the
time required to compute the OLS estimates. For both the jackknife and
score methods, there is considerable overlap between the computations
needed for the OLS estimates and the ones needed for CV$_{\tn3}$.
Thus, for large clusters, the cost of computing the OLS estimates and
CV$_{\tn3}$ together using one or both of these methods was sometimes
less than the cost of computing the OLS estimates alone. This is
probably because of cache congestion, which seems to be alleviated by
forming $\biX^\top\biX$ on a cluster-by-cluster basis. For large
clusters, the speed of all methods could almost certainly be increased
by using a fast BLAS implementation. However, in the interest of
programming ease, we have not done this. The jackknife and score
methods are already very fast.

In \Cref{fig:1}, the residual method works well for very small values
of $M$\tn. It is always the fastest method for $M\le4$. We did not
perform any timings for $M=1$, where CV$_{\tn3}$ reduces to HC$_3$,
because we would have needed a different program that eliminated the
loops within each cluster to obtain optimal results. But the residual
method is certainly the fastest one for this case. However, its cost
rises very rapidly as $M$ increases. Results for this method are only
shown for $M \le 4096$, because using it for larger values would have
been prohibitively costly. For the largest values of $M$\tn, the cost
of the residual method is almost the same for $k=20$ and $k=40$,
because it is dominated by the computations needed to form and invert
the $\biM_{gg}$ matrices.

In contrast, both the score and jackknife methods become faster as $M$
increases and $G$ consequently decreases, except that, when $k=40$,
they are both a bit slower for $M=64K$ than for $M=32K$. This probably
occurs because of cache congestion. The jackknife method is always
quicker than the score method. For small values of $M$\tn, it seems to
be faster by a factor of about 12 when $k=20$ and by a factor of about
26 when $k=40$. However, the advantage of the jackknife method
gradually diminishes as $M$ increases. When $M=64K$, so that there are
only 16 clusters, the jackknife method is only slightly faster.

It is easy to see that the jackknife method will have a big advantage
over the residual method whenever cluster sizes vary much, even if
most of them are very small. Imagine a sample with, say, 1000
equal-sized clusters and $M=5$. For such a sample, the residual and
jackknife methods will perform about the same. Suppose we then merge
100 of the tiny clusters into one large cluster with 500 observations.
Doing this will reduce the cost of the jackknife method slightly, but
it will greatly increase the cost of the residual method. Indeed, when
there is even a single very large cluster, the latter inevitably
becomes extremely slow.

Based on these results, the jackknife method for computing 
CV$_{\tn3}$ is clearly the procedure of choice unless all clusters are
tiny (say, $N_g \leq 5$ for all~$g$). For datasets with large
clusters, an efficient implementation of this method (such as the one
provided by the \texttt{summclust} package mentioned in
\Cref{sec:jack}), can compute both the OLS estimates and the
CV$_{\tn3}$ variance matrix in roughly the same amount of time as a
reasonably fast program for the OLS estimates alone.

\section{New Versions of the Wild Cluster Bootstrap}
\label{sec:boot}

The existing WCR bootstrap is based on CV$_{\tn1}$ standard errors and
the restricted empirical score vectors defined in \eqref{rescore}
below. Henceforth, we will refer to this as the classic WCR bootstrap,
or \mbox{WCR-C}. It often works well, but not always. We therefore
propose three new versions of the WCR bootstrap, along with three
corresponding versions of the WCU bootstrap. These are based on two
distinct modifications. One involves replacing CV$_{\tn1}$ by
CV$_{\tn3}$. The other involves modifying the scores used in the
bootstrap DGP, in the hope that the modified bootstrap DGP will
provide a better approximation to the unknown process that actually
generated the data.

We first discuss the bootstrap DGPs for all versions of the wild
cluster bootstrap, expressing them in terms of scores instead of
observations. This approach is intuitive and computationally
attractive \citep{RMNW,JGM-fast}. In terms of the $G$ score vectors, a
generic wild cluster bootstrap DGP is
\begin{equation}
\label{eq:wcscore}
\bis_g^{*b} = v_g^{*b} \ddot\bis_g, \quad g=1,\ldots,G,\quad
b=1,\ldots,B,
\end{equation}
where $b$ indexes bootstrap samples, $v_g^{*b}$ is a random variate
with mean~0 and variance~1, and the $\ddot\bis_g$ are empirical score
vectors to be discussed below. In most cases, it seems to be best to
generate the $v_g^{*b}$ using the Rademacher distribution, which takes
the values 1 and $-1$ with equal probabilities
\citep{DF_2008,DMN_2019}. However, since the number of possible
Rademacher bootstrap samples that are distinct from the original sample
is only $2^G - 1$, it is better to use a distribution with more mass 
points, such as the six-point distribution proposed in 
\citet{Webb_2022}, when $G$ is less than about~12.

The vector $\ddot\bis_g$ in \eqref{eq:wcscore} is an empirical score
vector for the \th{g} cluster. For the \mbox{WCU-C} bootstrap, it is
simply the unrestricted empirical score vector
$\hat\bis_g=\biX_g^\top\hat\biu_g$. For the \mbox{WCR-C} bootstrap, it
is the restricted empirical score vector $\tilde\bis_g$ defined as
\begin{equation}
\label{rescore}
\tilde\bis_g = \biX_g^\top\biy_g - \biX_g^\top\biX_g\tilde\bbeta,\quad
g=1,\ldots,G,
\end{equation}
where $\tilde\bbeta$ is the vector of OLS estimates under the null
hypothesis. Like $\hat\bbeta$, $\tilde\bis_g$ is a $k$-vector, even
though some elements of $\tilde\bbeta$ may equal zero or satisfy other
linear restrictions. The bootstrap DGP \eqref{eq:wcscore} looks very
much like the one for the wild score cluster bootstrap for nonlinear
models proposed in \citet{KS_2012a}. In the context of 
\eqref{eq:lrmodel}, however, it is just a different way of writing the
bootstrap DGP for the wild cluster bootstrap.

In order to calculate a bootstrap $P$ value or a bootstrap confidence
interval, we need to compute $B$ bootstrap test statistics indexed
by~$b$. These depend only on the bootstrap scores in
\eqref{eq:wcscore} and the matrix $(\biX^\top\biX)^{-1}$\tn.  For each
bootstrap sample, we use $\bis_g^{*b}$ to obtain a bootstrap estimate,
not of $\bbeta$ itself, but of the vector $\bdelta = \bbeta -
\ddot\bbeta$, where $\ddot\bbeta = \tilde\bbeta$ for the \mbox{WCR}
bootstrap and $\ddot\bbeta = \hat\bbeta$ for the \mbox{WCU}
bootstrap. This estimate is simply
\begin{equation}
\label{bootbeta}
\hat\bdelta^{*b} = (\biX^\top\biX)^{-1} \sum_{g=1}^G\bis_g^{*b} =
(\biX^\top\biX)^{-1} \bis^{*b},
\end{equation}
where $\bis^{*b} = \sum_{g=1}^G\bis_g^{*b}$. When $v_g^{*b}=1$ for all
$g$, the bootstrap sample is the same as the original sample. In this
very special case, $\hat\bdelta^{*b} = \bzero$ for the \mbox{WCU}
bootstrap, and $\hat\bdelta^{*b} = \hat\bbeta - \tilde\bbeta$ for the
\mbox{WCR} bootstrap.

If we are testing the hypothesis that $\beta_j=0$, where $\beta_j$ is
an element of $\bbeta$, then we just need to multiply the \th{j} row
of $(\biX^\top\biX)^{-1}$ by $\bis^{*b}$ in order to obtain
$\hat\delta_j^{*b}$, the \th{j} element of $\bdelta^{*b}$\tn. The
bootstrap $t$-statistic is then equal to
\begin{equation}
\label{boottj}
t_j^{*b} = \frac{\hat\delta_j^{*b}}{{\rm se}(\hat\delta_j^{*b})}\tk,
\end{equation}
where ${\rm se}(\cdot)$ denotes the standard error formula used to
obtain $t_j$, the original $t$-statistic. We automatically get the
correct numerator, which is $\hat\beta_j^{*b}$ for the \mbox{WCR}
bootstrap, since $\ddot\bbeta=\tilde\bbeta$, and $\hat\beta_j^{*b} -
\hat\beta_j$ for the \mbox{WCU} bootstrap, since
$\ddot\bbeta=\hat\bbeta$. As usual, a symmetric bootstrap $P$ value is
then given by
\begin{equation}
P_{\rm S}^*(t_j) = \frac1B\sum \IF\big(|t_j^{*b}| > |t_j|\big),
\label{eq:bootp}
\end{equation}
where $\IF(\cdot)$ denotes the indicator function. The bootstrap $P$
value in \eqref{eq:bootp} is simply the fraction of the bootstrap
samples for which $|t_j^{*b}|$ is more extreme than $|t_j|$. The value
of $B$ should be chosen so that $\alpha(B+1)$ is an integer, where
$\alpha$ is the level of the test \citep{RM_2007}. It is common to use
$B=999$, but $B=9,\tn999$ and (when feasible) $B=99,\tn999$ are better
choices.

In the classic versions of the wild cluster bootstrap, the standard
error formula in \eqref{boottj} is ${\rm se}_1(\cdot)$, the square 
root of the \th{j} diagonal element of CV$_{\tn1}$. But the results in
\Cref{sec:jack} make it equally feasible to use standard errors based 
on CV$_{\tn3}$, even in large samples. This gives us new versions of
both the WCR and WCU bootstraps, which we will refer to as
\mbox{WCR-V} and \mbox{WCU-V}, because only the variance matrices have
changed. The bootstrap standard errors can be calculated without
computing an entire variance matrix for each bootstrap sample. For
example, the CV$_{\tn3}$ standard error of $\hat\delta_j^{*b}$ is just
\begin{equation}
{\rm se}_3(\hat\delta_j^{*b}) = 
\left(\frac{G-1}{G} \sum_{g=1}^G \big(\hat\delta_{j(g)}^{*b}
- \hat\delta_j^{*b}\big)^2\right)^{\!\!1/2}\tn,
\label{eq:jackse3}
\end{equation}
where $\hat\delta_{j(g)}^{*b}$ is the \th{j} element of the vector
\begin{equation}
\label{eq:deloneboot}
\hat\bdelta_{(g)}^{*b} = (\biX^\top\!\biX - \biX_g^\top\!\biX_g)^{-1}
(\bis^{*b} - \bis_g^{*b}).
\end{equation}
Only $\hat\delta_j^{*b}$ and the $\hat\delta_{j(g)}^{*b}$ need to be
computed for each bootstrap sample. In \eqref{bootbeta} and 
\eqref{eq:deloneboot}, the first terms are invariant across bootstrap 
samples and only need to be computed once.

We now have two versions of the WCR bootstrap, \mbox{WCR-C} and
\mbox{WCR-V}, and two versions of the WCU bootstrap, \mbox{WCU-C} and
\mbox{WCR-V}. The two WCR bootstraps use the bootstrap DGP
\eqref{eq:wcscore} with $\ddot\bis_g = \tilde\bis_g$, and the two WCU
bootstraps use the bootstrap DGP \eqref{eq:wcscore} with $\ddot\bis_g
= \hat\bis_g$. The ``C'' and ``V'' versions calculate both the actual
and bootstrap test statistics using ${\rm se}_1(\cdot)$ and ${\rm
se}_3(\cdot)$, respectively. These bootstrap methods use the
restricted or unrestricted empirical scores in their raw form. But
empirical scores differ from true scores, because residuals differ
from disturbances. It therefore seems attractive to replace the
empirical score vectors by modified score vectors that implicitly
rescale the residuals on a cluster-by-cluster basis. This is analogous
to methods discussed in \citet{DF_2008} and \citet{JGM_2013} for the
ordinary wild bootstrap. However, quite a lot more algebra is needed.

For the WCU bootstrap, we can simply replace the vectors $\ddot\bis_g$
in \eqref{eq:wcscore} with the modified empirical score vectors
$\acute\bis_g$ defined in \eqref{eq:bis3}. Using \eqref{eq:bis3} is
expensive for large clusters, but the result \eqref{eq:keyres} lets us
compute $\acute\bis_g$ very rapidly as
\begin{equation}
\acute\bis_g = \biX^\top\biX\big(\hat\bbeta - \hat\bbeta^{(g)}\big), \quad
g=1,\ldots,G.
\label{eq:jackequal}
\end{equation}
For large clusters, using \eqref{eq:bis3f} to compute the
$\acute\bis_g$ is much faster than using \eqref{eq:bis3}, but using
\eqref{eq:jackequal} is faster still; see \Cref{sec:speed}. This
yields two new bootstrap methods, which we will refer to as
\mbox{WCU-S} and \mbox{WCU-B}, respectively. The \mbox{WCU-S}
bootstrap (S for score) employs the modified score vectors
$\acute\bis_g$ instead of $\hat\bis_g$, but it uses the familiar ${\rm
se}_1(\cdot)$ standard error. The \mbox{WCU-B} bootstrap (B for both)
employs both the modified score vectors and the ${\rm se}_3(\cdot)$
standard error.

Finding the analogous versions of the WCR bootstrap takes a bit more
work. We need to specify a restricted wild bootstrap DGP based on
modified score vectors. Suppose the restrictions have the usual linear
form, $\biR \bbeta = \bir$, for a given matrix $\biR$ and a given
vector $\bir$. We can write this equivalently in terms of free
parameters, $\bm\phi$, as $\bbeta = \biH\bm\phi +\bih$ for a given
matrix $\biH$ and a given vector~$\bih$. Then the modified score
vectors are
\begin{equation}
\dot\bis_g =\biX_g^\top\tilde\biM_{gg}^{-1}(\biy_g -\biX_g\tilde\bbeta ),
\label{eq:bisr3}
\end{equation}
which are the analogs of the $\acute\bis_g$ from \eqref{eq:bis3}. Here
$\tilde\biM_{gg}$ is the \th{g} diagonal block of the projection matrix
$\tilde\biM = \bfI -\tilde\biX (\tilde\biX^\top\tilde\biX )^{-1}
\tilde\biX^\top$, where $\tilde\biX = \biX\biH$. However, evaluating 
\eqref{eq:bisr3} is computationally infeasible when the clusters are 
not all small. We need to replace \eqref{eq:bisr3} by something that is 
feasible for any sample size.

The first step is to compute $\tilde\bbeta = \biH\tilde{\bm\phi}
+\bih$, where $\tilde\biy = \biy - \biX\bih$ and $\tilde{\bm\phi} = 
(\tilde\biX^\top\tilde\biX)^{-1}\tilde\biX^\top\tilde\biy$. The
corresponding estimates when cluster $g$ is omitted are
$\tilde\bbeta^{(g)} = \biH\tilde{\bm\phi}^{(g)} +\bih$, where
\begin{equation}
\label{eq:deloner}
\tilde{\bm\phi}^{(g)} = (\tilde\biX^\top\tilde\biX - \tilde\biX_g^\top
\tilde\biX_g )^{-1} (\tilde\biX^\top\tilde\biy - \tilde\biX_g^\top
\tilde\biy_g ), \quad g=1,\ldots,G.
\end{equation}
Then it can be shown that
\begin{equation}
\dot\bis_g = \biX_g^\top\tkk\tilde\biy_g -
\biX_g^\top\tilde\biX_g\tilde{\bm\phi}^{(g)}, \quad g=1,\ldots,G.
\label{eq:jackeqr}
\end{equation}
To see that \eqref{eq:bisr3} and \eqref{eq:jackeqr} are equal, note that
the right-hand side of \eqref{eq:jackeqr} is
\begin{align*}
&\biX_g^\top \big( \tilde\biy_g - \tilde\biX_g (\tilde\biX^\top\tilde\biX 
- \tilde\biX_g^\top\tilde\biX_g)^{-1}(\tilde\biX^\top\tilde\biy 
- \tilde\biX_g^\top\tilde\biy_g ) \big) \\
&\quad = \biX_g^\top \big( \tilde\biy_g - \tilde\biX_g \big(
(\tilde\biX^\top\tilde\biX)^{-1} + (\tilde\biX^\top\tilde\biX)^{-1} \tilde\biX_g^\top 
\tilde\biM_{gg}^{-1}\tilde\biX_g (\tilde\biX^\top\tilde\biX)^{-1} \big)
(\tilde\biX^\top\tilde\biy - \tilde\biX_g^\top\tilde\biy_g ) \big) ,
\end{align*}
where the equality uses the updating formula \eqref{eq:update} applied
to $\tilde\biX$\tn, $\tilde\biX_g$, and $\tilde\biM_{gg}^{-1}$. Then we
use the fact that $\tilde{\bm\phi} =
(\tilde\biX^\top\tilde\biX)^{-1}\tilde\biX^\top \tilde\biy$ together
with the relation $\tilde\biX_g (\tilde\biX^\top
\tilde\biX)^{-1}\tilde\biX_g^\top = \tilde\biP_{gg} =
\bfI-\tilde\biM_{gg}$ to rewrite the last expression as
\begin{equation}
\begin{aligned}
&\biX_g^\top \big( \tilde\biy_g - \tilde\biX_g \tilde{\bm\phi}
-(\bfI-\tilde\biM_{gg})\tilde\biM_{gg}^{-1}\tilde\biX_g\tilde{\bm\phi}
+(\bfI-\tilde\biM_{gg})\tilde\biy_g
+(\bfI-\tilde\biM_{gg})\tilde\biM_{gg}^{-1}
(\bfI-\tilde\biM_{gg})\tilde\biy_g \big) \\
&\quad = \biX_g^\top \tilde\biM_{gg}^{-1} ( \tilde\biy_g - \tilde\biX_g 
\tilde{\bm\phi} ) .
\end{aligned}
\label{eq:intexp}
\end{equation}
Replacing $\tilde\biy_g$ by $\biy_g -\biX_g \bih$ and $\tilde\biX_g$ by 
$\biX_g \biH$, and using the fact that $\biH\tilde{\bm\phi} = 
\tilde\bbeta -\bih$, the right-hand side of \eqref{eq:intexp} 
equals~\eqref{eq:bisr3}.

An important special case is the restriction that $\beta_k = 0$. This
is obtained by setting $\biR = ( 0,\ldots ,0,1 )$ and $\bir =0$, or,
equivalently, $\biH = (\bfI_{k-1}, \bzero)^\top$ and $\bih = \bzero$.
In this case, we find that $\tilde\biX = \biX_1$, which contains the
first $k-1$ columns of $\biX$\tn, and $\tilde{\bm\phi}=\tilde\bbeta_1
= (\biX_1^\top\biX_1)^{-1}\biX_1^\top\biy$. The corresponding
estimates when each cluster is omitted in turn are
\begin{equation}
\label{eq:delrest}
\tilde\bbeta_1^{(g)} = (\biX_1^\top\!\biX_1 - \biX_{1g}^\top\tk\biX_{1g})^{-1}
(\biX_1^\top\biy - \biX_{1g}^\top\tk\biy_g), \quad g=1,\ldots,G,
\end{equation}
where $\biX_{1g}$ contains the first $k-1$ columns of $\biX_g$. Then 
\eqref{eq:jackeqr} reduces to
\begin{equation}
\dot\bis_g = \biX_g^\top\tkk\biy_g -
\biX_g^\top\tk\biX_{1g}\tkk\tilde\bbeta_1^{(g)}, \quad g=1,\ldots,G.
\label{eq:jacksimple}
\end{equation}

Exactly the same arguments that led to \eqref{eq:jackeqr} can be
applied to the modified unrestricted empirical scores, giving us
\begin{equation}
\label{eq:jackequ}
\acute\bis_g = \biX_g^\top\tkk\biy_g -
\biX_g^\top\biX_g\hat\bbeta^{(g)}, \quad g=1,\ldots,G.
\end{equation}
Either \eqref{eq:jackequal} or \eqref{eq:jackequ} can be used to
compute the $\acute\bis_g$, and both are computationally attractive.
However, in situations where both $\dot\bis_g$ and $\acute\bis_g$ need
to be computed, \eqref{eq:jackequ} may offer some programming
advantages relative to \eqref{eq:jackequal} due to its similarity to
\eqref{eq:jackeqr}.

The scalar factors in \eqref{eq:CV1} and \eqref{eq:CV3} do not appear
in the bootstrap DGPs that correspond to them, because rescaling all
the bootstrap scores by the same factor has no impact on the 
bootstrap $t$-statistics. From \eqref{bootbeta} and
\eqref{eq:deloneboot}, multiplying all the $\bis_g^{*b}$ by a scalar
$C$ simply makes $\hat\bdelta^{*b}$ and all the
$\hat\bdelta_{(g)}^{*b}$ larger by a factor of~$C$. This also makes
the empirical scores for every bootstrap sample larger by the same
factor. Therefore, from \eqref{eq:CV1}, \eqref{eq:CV2}, and
\eqref{eq:CV3}, the variance matrices become larger by a factor of
$C^{\tkk2}$ and the standard errors by a factor of~$C$. The factors of
$C$ in the numerator and denominator of $t_j^{*b}$ cancel out, leaving
the bootstrap $t$-statistics unchanged.

However, no cancellation occurs for bootstrap tests of $\beta_j=0$
based directly on $\hat\beta_j$ and its bootstrap analog of
$\hat\delta_j^{*b}$. In this case, multiplying the right-hand side of
\eqref{eq:wcscore} by the square root of $G(N-1)/((G-1)(N-k))$ for
methods that use CV$_{\tn1}$ and by the square root of $(G-1)/G$ for
methods that use CV$_{\tn3}$ should improve the correspondence between
the bootstrap DGP and the unknown true DGP. The usual theory of
higher-order refinements for the bootstrap suggests that it is
generally better to studentize \citep{Hall_1992}. However, there may
be cases in which unstudentized test statistics are of interest
\citep*{CSS_2021}. But since we have eight studentized bootstrap
methods to study, we do not consider unstudentized ones further.

To generate the transformed scores needed for the \mbox{WCR/WCU-S} and
\mbox{WCR/WCU-B} bootstraps, \eqref{eq:jackequal} and
\eqref{eq:jackequ} must be used for all $G$ clusters. In the event
that $\hat\bbeta^{(h)}$ and $\tilde\bbeta^{(h)}$ cannot be calculated
for cluster~$h$, we have two choices. The simplest is to replace the
inverses in \eqref{eq:delone} and \eqref{eq:delrest} by generalized
inverses. Alternatively, we could use $\hat\bis_h$ instead of
$\acute\bis_h$ and $\tilde\bis_h$ instead of $\dot\bis_h$, along with
the transformed scores for the remaining clusters. The latter would be
appropriate if we have chosen to omit the problematic clusters when
computing the cluster-jackknife variance matrix; see the discussion at
the end of \Cref{sec:jack}.

\begin{table}[tp]
\caption{Eight versions of the wild cluster bootstrap}
\label{tab:eight}
\vspace*{-2em}
\begin{center}
\begin{tabular*}{0.8\textwidth}{@{\extracolsep{\fill}}lcc}
\toprule
&\multicolumn{2}{c}{Standard errors based on}\\
\cmidrule{2-3}
Scores in bootstrap DGP \eqref{eq:wcscore} &CV$_{\tn1}$ &CV$_{\tn3}$ \\
\midrule
Null hypothesis imposed & & \\
\addlinespace[0.5ex]
$\tilde\bis_g$ defined in \eqref{rescore} &WCR-C &WCR-V \\
$\dot\bis_g$ defined in \eqref{eq:jacksimple} &WCR-S &WCR-B \\
\midrule
Null hypothesis not imposed & & \\
\addlinespace[0.5ex]
$\hat\bis_g = \biX_g^\top \hat\biu_g$ &WCU-C &WCU-V \\
$\acute\bis_g$ defined in \eqref{eq:jackequal} or \eqref{eq:jackequ}
&WCU-S &WCU-B \\
\bottomrule
\end{tabular*}
\vskip 4pt \parbox{0.8\textwidth}
{\footnotesize
\textbf {Notes:} WCR-C and WCU-C are the classic versions of the wild
cluster restricted and wild cluster unrestricted bootstraps.
\mbox{WCR-S} and \mbox{WCU-S} employ transformed scores with the usual
CV$_{\tn1}$ variance matrix. \mbox{WCR-V} and \mbox{WCU-V} employ the
usual scores with the CV$_{\tn3}$ variance matrix. \mbox{WCR-B} and
\mbox{WCU-B} employ both transformed scores and CV$_{\tn3}$.}
\end{center}
\end{table}

\Cref{tab:eight} provides a convenient summary of the eight wild
cluster bootstrap methods that we have discussed. Conceptually, they
differ along two dimensions. The horizontal dimension represents the
way in which the standard errors for both the actual and bootstrap
test statistics are calculated. The vertical dimension represents the
score vectors used in the four versions of the bootstrap DGP
\eqref{eq:wcscore}. Note that the \texttt{boottest} and
\texttt{fwildclusterboot} packages now provide fast implementations of
the \mbox{WCR/WCU-S} bootstraps as well as the classic ones. This is
possible because, in contrast to the \mbox{WCR/WCU-V} and
\mbox{WCR/WCU-B} bootstraps, the former do not involve any jackknife
calculations for the bootstrap samples. Once the transformed scores
have been computed, the fast bootstrap algorithm proposed in
\citet{RMNW} applies directly to the \mbox{WCR/WCU-S} bootstraps.

It seems highly likely that all the methods discussed in this section
are asymptotically valid, in the sense that, under suitable regularity
conditions, the rejection frequencies for any test converge to the
nominal level of the test as $G\to\infty$. Formal proofs could be
obtained by modifying the arguments in \citet{DMN_2019}. For the WCU
bootstrap methods, the key fact is that the modified empirical score
vectors $\acute\bis_g$ computed using \eqref{eq:jackequal} or
\eqref{eq:jackequ} are asymptotically equal to the ordinary empirical
score vectors $\hat\bis_g$. For the WCR bootstrap methods, the key
fact is that the modified restricted empirical score vectors
$\dot\bis_g$ defined in \eqref{eq:jackeqr} are asymptotically equal to
the ordinary restricted empirical score vectors $\tilde\bis_g$
in~\eqref{rescore}.

\section{Monte Carlo Simulations}
\label{sec:sims}

Simulation results in \citet{MW-JAE,MW-EJ}, \citet{Brewer_2018},
\citet{DMN_2019}, \citet{JGM-fast}, and several other papers have
shown that the reliability of bootstrap and asymptotic methods for
cluster-robust inference depends heavily on the number of clusters,
the extent to which cluster sizes vary, and (in the case of treatment
effects) both the number of treated clusters and their sizes. Many of
our experiments therefore focus on these features.

The model we consider is
\begin{equation}
y_{gi} = \beta_1 + \sum_{j=2}^k \beta_j X_{jgi} + u_{gi}, \quad
g=1,\ldots,G, \quad i=1,\ldots,N_g,
\label{eq:simmod}
\end{equation}
where the $u_{gi}$ are generated by a normal random-effects model with
intra-cluster correlation~$\rho$. The way in which the $k-1$
non-constant regressors are generated varies across the experiments.
The hypothesis to be tested is that~$\beta_k=0$.

In most of our experiments, there are $N=400G$ observations, which are
divided among the $G$ clusters using the formula
\begin{equation}
N_g = \left[N \frac{\exp(\gamma g/G)}{\sum_{j=1}^G \exp(\gamma
j/G)}\right]\!, \quad g=1,\ldots, G-1,
\label{eq:gameq}
\end{equation}
where $[x]$ means the integer part of $x$. The value of $N_G$ is then
set to $N - \sum_{g=1}^{G-1} N_g$. The key parameter here is $\gamma$,
which determines how uneven the cluster sizes are. When $\gamma=0$ and
$N/G$ is an integer, \eqref{eq:gameq} implies that $N_g = N/G$ for 
all~$g$. For $\gamma>0$, cluster sizes vary more and more as $\gamma$ 
increases. The largest value of $\gamma$ that we use is~4. In that 
case, when $G=24$ and $N=9600$, the largest cluster (1513
observations) is about 47 times as large as the smallest cluster (32
observations). In contrast, when $\gamma=2$, the largest cluster (899
observations) is just under seven times as large as the smallest (130
observations).

The sample sizes that we employ are unusually large for experiments of
this type. Since cluster-robust inference is often used with samples
that have hundreds of thousands or even millions of observations, we
want our results to apply to such cases. In preliminary experiments,
we found that the results tended to change slightly, but 
systematically, as small values of $N/G$ were increased. Results for
$N/G>400$ are very similar to ones for $N/G=400$, so we use 400 in all
the experiments based on \eqref{eq:gameq}. Because the bootstrap
samples are generated using scores, the cost of the experiments
increases much less than proportionally with $N/G$.

All experiments use $400,\tn000$ replications. This number is so large
that experimental randomness is negligible. The most important
determinant of computational cost is $k$, the number of regressors. As
can be seen from \eqref{eq:wcscore} and \eqref{eq:jackeqr} or
\eqref{eq:jackequ}, generating each bootstrap sample involves
$O(k^2G)$ operations. So does calculating the test statistics using
either CV$_{\tn1}$ or CV$_{\tn3}$. Thus the experiments can be
somewhat costly when $k$ is large.\footnote{The method of
\citet{RMNW}, which can only be used for the WCR/WCU-C and WCR/WCU-S
bootstraps, is usually less expensive when $k$ is not small, but our
programs did not use it.} Nevertheless, many of our experiments
involve $k\ge10$. We do this because results in \citet{JGM-fast}
suggest that the performance of many methods of inference deteriorates
as $k$ increases. Previous Monte Carlo experiments, which often use
$k\le3$, may therefore have tended to give too optimistic a picture.

It might seem that substantial savings could be achieved by partialing
out all regressors except the one(s) of interest prior to performing
the bootstrap. However, this trick only works in certain special
cases. For methods based on the jackknife, it is easy to see the
problem. If we were to partial out some of the regressors prior to
computing the delete\tkk-one\tkk-cluster estimates in
\eqref{eq:delone}, then the computed $\hat\bbeta^{(g)}$ would depend
on the values of the partialed-out regressors for the full sample,
including those in the \th{g} cluster which was supposed to be
deleted. Consequently, the values of the delete\tkk-one\tkk-cluster
estimates would be incorrect if we partialed out any regressor that
affects more than one cluster (such as industry-level fixed effects
with firm-level clustering).

An important exception is when the regressors that are partialed out
are cluster fixed effects or fixed effects at a finer level (such as
firm-level fixed effects with industry-level clustering), because each
of them affects only some or all of the observations within a single
cluster. In fact, it is essential to partial out fixed effects of this
type if using a generalized inverse is to be avoided.

\subsection{Test Size}
\label{subsec:size}

\begin{figure}[tb]
\begin{center}
\caption{Rejection frequencies as a function of $\gamma$}
\vspace*{-0.5em}
\label{fig:2}
\includegraphics[width=0.95\textwidth]{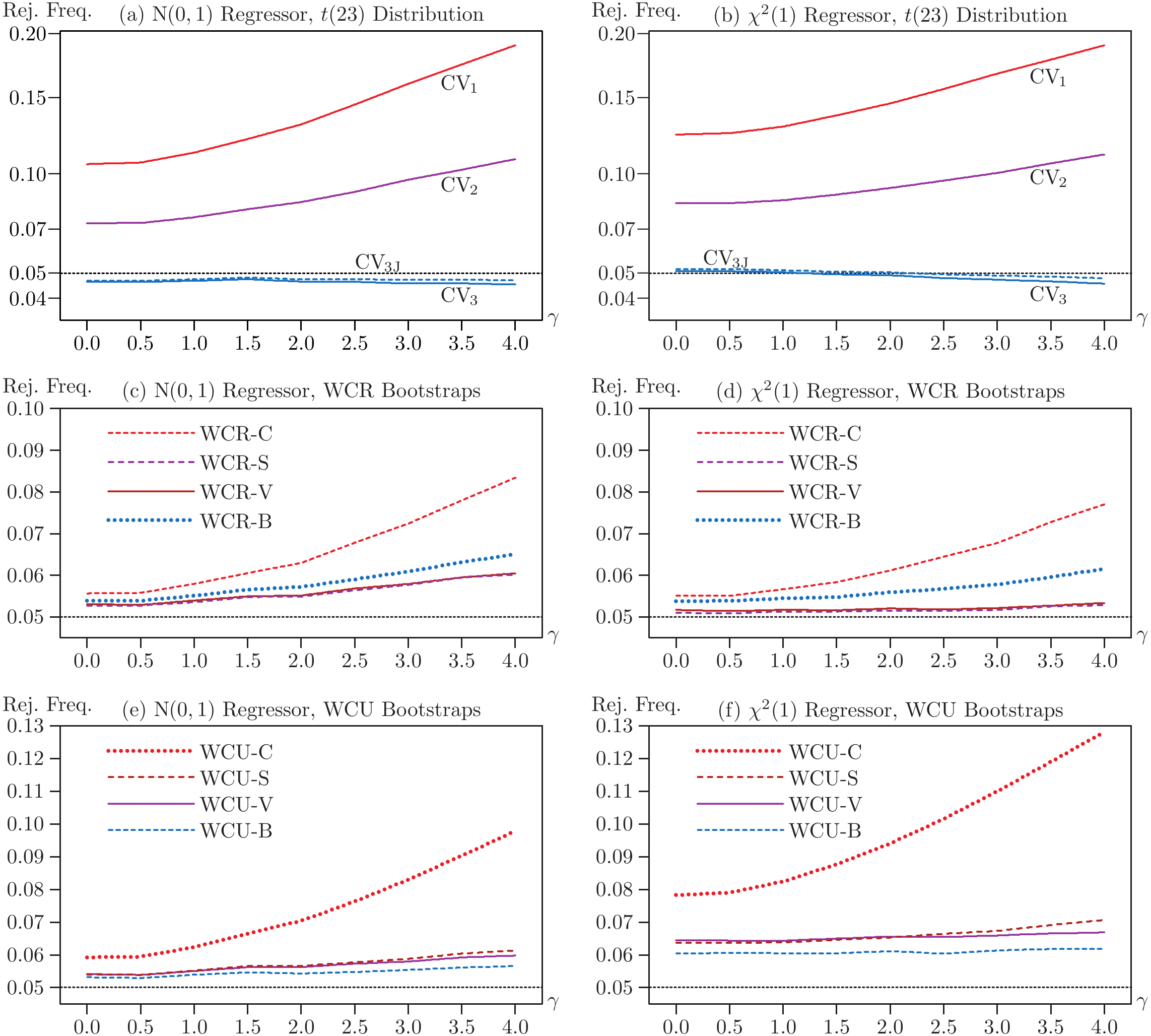}
\end{center}
{\footnotesize \textbf{Notes:} The vertical axes show rejection
frequencies for tests of $\beta_k=0$ in \eqref{eq:simmod} 
at the .05 level. Results are based on $400,\tn000$ replications, with
$B=399$ bootstrap samples. There are 24 clusters, 9600 observations,
and 10 regressors, with $\rho=0.10$. The extent to which cluster sizes
vary increases with $\gamma$; see \eqref{eq:gameq}.}
\end{figure}

The experiments in this subsection deal with rejection frequencies
under the null hypothesis. We consider both asymptotic tests based on
the $t(G-1)$ distribution and the wild cluster bootstrap tests listed
in \Cref{tab:eight}.

The experiments in \Cref{fig:2} focus on variation in cluster sizes.
There are always 9600 observations, 24 clusters, and 10 regressors.
Cluster sizes vary according to \eqref{eq:gameq}. Regressors 2 through
$k-1$ in \eqref{eq:simmod} follow a normal random-effects model that
yields intra-cluster correlations of~0.50. The test regressor either
follows the same normal distribution as the others (in the three
panels on the left), or a $\chi^2(1)$ distribution (in the three
panels on the right). In the latter case, it is the square of a
normally distributed random variable generated by the same
random-effects model as the other regressors. The disturbances are
also generated by such a model, but with $\rho=0.10$. We focus on
rejection frequencies for a test that $\beta_k=0$ at the 5\% level.

The results for asymptotic tests, based on the $t(23)$ distribution
and shown in Panels~(a) and~(b), are striking. Note that a
square\tkk-root transformation has been applied to the vertical axis
to prevent these panels from being too tall. Tests based on
CV$_{\tn1}$ over-reject substantially. The extent of the
over-rejection increases with $\gamma$, and, except for $\gamma=4$, it
is more severe in Panel~(b) than in Panel~(a). A regressor that
follows the $\chi^2(1)$ distribution necessarily has some extreme
values. These become points of high leverage, which makes inference
more difficult in Panel~(b).

Although tests based on CV$_{\tn2}$ always reject considerably less
often than ones based on CV$_{\tn1}$, they also over-reject
significantly and to an extent that increases with $\gamma$. In
contrast, tests based on CV$_{\tn3}$ and CV$_{\tn3{\rm J}}$ either
under-reject slightly all the time, in Panel~(a), or under-reject very
slightly for larger values of~$\gamma$, in Panel~(b). The results for
CV$_{\tn3}$ and CV$_{\tn3{\rm J}}$ are extremely similar. The latter
always rejects more often than the former, because the difference
between \eqref{eq:jack} and \eqref{eq:jack3} is the positive
semi-definite matrix $((G-1)/G)
(\hat\bbeta-\bar\bbeta)(\hat\bbeta-\bar\bbeta)^\top$. Since
CV$_{\tn3}$ tends to under-reject slightly in \Cref{fig:2}, it might 
seem that CV$_{\tn3{\rm J}}$ is to be preferred. However, as we shall 
see, there are also many cases in which CV$_{\tn3}$ over-rejects, and
CV$_{\tn3{\rm J}}$ therefore over-rejects slightly more. In
practice, it would be perfectly reasonable to report either
CV$_{\tn3}$ or CV$_{\tn3{\rm J}}$. We never encountered a case in
which it made any real difference.

The results for the WCR bootstrap tests, shown in Panels (c) and~(d),
are surprising. In the past, \mbox{WCR-C} has been the only variant of
the WCR bootstrap, and numerous Monte Carlo experiments have suggested
that it is the procedure of choice. But \mbox{WCR-B} performs notably
better than \mbox{WCR-C} for every value of $\gamma$, and both
\mbox{WCR-V} and \mbox{WCR-S} perform better still. Note that,
although these two procedures perform almost the same here, this is
not true in general. Oddly, all the WCR procedures perform better in
Panel~(d), where the test regressor is highly skewed, than they do in
Panel~(c), where it is Gaussian. The rather mediocre performance of
\mbox{WCR-C} must be due, at least in part, to the fact that $k=10$,
which is a larger number than has been used in most previous
experiments; see \Cref{fig:3} below.

\begin{figure}[tb]
\begin{center}
\caption{Rejection frequencies as a function of $k$}
\vspace*{-0.5em}
\label{fig:3}
\includegraphics[width=0.95\textwidth]{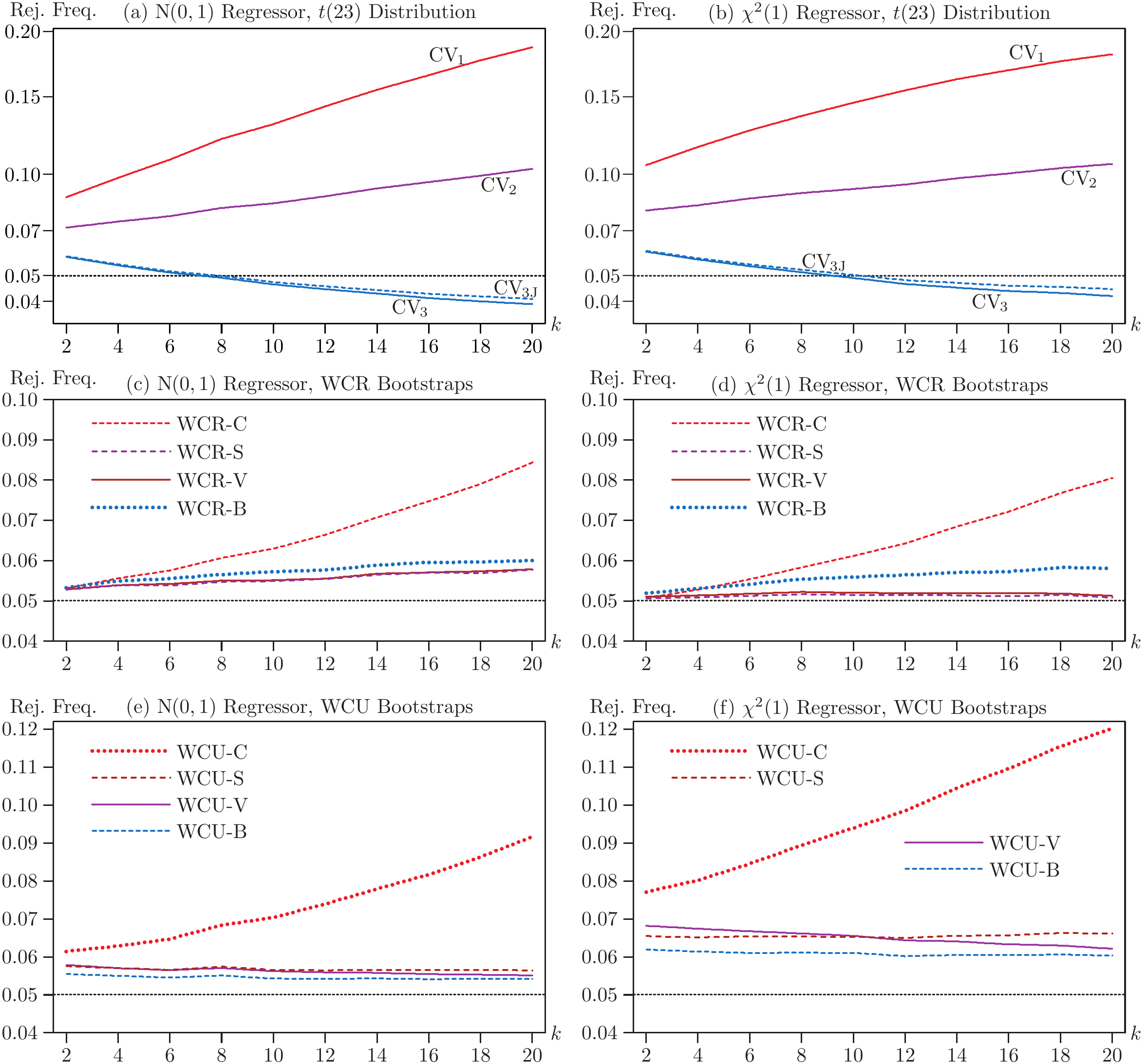}
\end{center}
{\footnotesize \textbf{Notes:} The vertical axes show rejection 
frequencies for tests of $\beta_k=0$ in \eqref{eq:simmod} at the .05 level.
Results are based on $400,\tn000$ replications, with $\gamma=2$,
$\rho=0.10$, and $B=399$ bootstrap samples. There are 24 clusters,
9600 observations, and $k$ regressors, where $k$ varies from 2 to 20
by~2.}
\end{figure}

Some of the results for the WCU bootstrap tests, shown in Panels (e)
and~(f), are also surprising. It is not a surprise that \mbox{WCU-C}
rejects more often than \mbox{WCR-C} or that its performance is much
worse in Panel~(f) than in Panel~(e). However, the fact that the other
three WCU procedures over-reject much less often than \mbox{WCU-C} may
well be surprising. In both panels, \mbox{WCU-B} is clearly the
procedure of choice. \mbox{WCU-V} and \mbox{WCU-S} perform much better
than \mbox{WCU-C}, but worse than \mbox{WCU-B}. In Panels (c) and~(d),
the differences between \mbox{WCU-V} and \mbox{WCU-S} are small, but
larger than the differences between \mbox{WCR-V} and \mbox{WCR-S}.

\Cref{fig:3} is similar to \Cref{fig:2}, but the number of regressors
$k$ is now on the horizontal axis, and $\gamma=2$. In Panels (a) 
and~(b), CV$_{\tn1}$ over-rejects to an increasing extent as $k$
increases. So does CV$_{\tn2}$, although it always over-rejects
considerably less than CV$_{\tn1}$. In contrast, CV$_{\tn3}$ and
CV$_{\tn3{\rm J}}$ over-reject modestly for small values of $k$ and
under-reject modestly for large ones.

Panels (c) and~(d) look a lot like the same panels in \Cref{fig:2},
even though what is on the horizontal axis is different. \mbox{WCR-C}
performs quite well for very small values of $k$, but it over-rejects 
more and more severely as $k$ increases. \mbox{WCR-B} performs much 
better than \mbox{WCR-C}, but \mbox{WCR-V} and \mbox{WCR-S} perform
even better. In Panel~(d), where the test regressor is highly skewed,
they both perform extremely well for all values of~$k$.

Panels (e) and~(f) also look a lot like the same panels in
\Cref{fig:2}. \mbox{WCU-C} performs quite poorly, over-rejecting more
and more severely as $k$ increases. In contrast, \mbox{WCU-B} performs
quite well in Panel~(e) and fairly well in Panel~(f), and there is no
tendency for its performance to deteriorate as $k$ increases. As
before, the two other bootstrap methods generally perform much better
than \mbox{WCU-C} but slightly worse than \mbox{WCU-B}.

\begin{figure}[tb]
\begin{center}
\caption{Rejection frequencies as a function of $G$}
\vspace*{-0.5em}
\label{fig:4}
\includegraphics[width=0.95\textwidth]{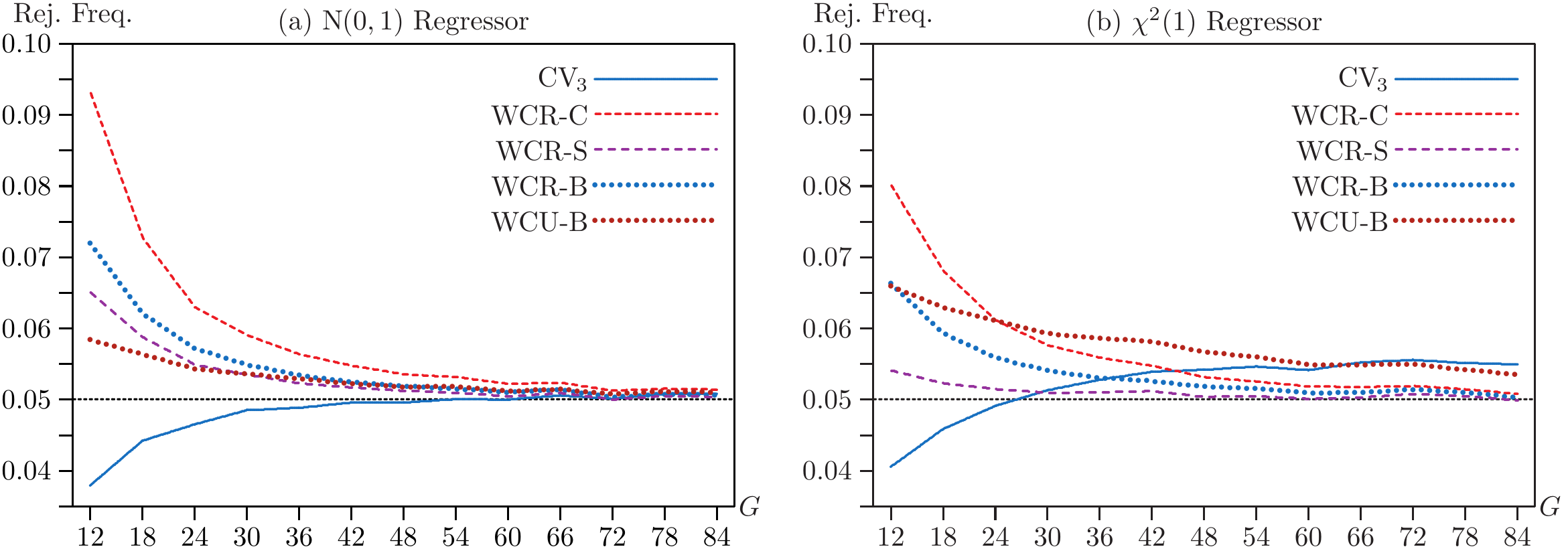}
\end{center}
{\footnotesize \textbf{Notes:} The vertical axes show rejection 
frequencies for tests of $\beta_k=0$ in \eqref{eq:simmod} at the .05
level. Results are based on $400,\tn000$ replications, with $\gamma=2$,
$k=10$, $\rho=0.10$, and $B=399$ bootstrap samples. There are between
12 and 84 clusters, all multiples of 6, with 400 observations per
cluster on average.}
\end{figure}

In the next set of experiments, we focus on what happens as $G$
increases. \Cref{fig:4} shows rejection frequencies as functions of
$G$, which varies from 12 to 84 by~6, and implicitly also $N$\tn,
since $N=400G$. In these experiments, $\gamma=2$ and $k=10$. We report
results for only five methods, instead of twelve. We omit CV$_{\tn1}$
and CV$_{\tn2}$, because they never perform very well, and
CV$_{\tn3{\rm J}}$ because it is almost identical to CV$_{\tn3}$.
Among the restricted bootstrap methods, we report \mbox{WCR-C},
because it was until now the procedure of choice. We also report
\mbox{WCR-S} and \mbox{WCR-B}, but we do not report \mbox{WCR-V},
because it yields results nearly identical to those of \mbox{WCR-S}
and is harder to compute. Among the unrestricted bootstrap methods, we
report only \mbox{WCU-B}, because it always seems to outperform the
other WCU methods.

In Panel~(a), using CV$_{\tn3}$ with the $t(G-1)$ distribution
under-rejects quite noticeably for very small values of~$G$, but it
performs extremely well for $G \geq 30$. The bootstrap methods always
over-reject, with \mbox{WCR-C} always the worst of them. For $G \geq 42$,
however, all the bootstrap methods perform very well, with \mbox{WCR-S}
the winner by a tiny margin.

Panel~(b) is more interesting than Panel~(a). The extreme skewness of
the $\chi^2(1)$ regressor apparently affects the results quite a bit,
even when $G=84$. Using CV$_{\tn3}$ with the $t(G-1)$ distribution
under-rejects for small values of $G$ but over-rejects for larger
values, where it is the worst method. Note that $G=24$, the value in
\Cref{fig:2,fig:3}, is near where the curve for CV$_{\tn3}$ crosses
the .05 line in \Cref{fig:4}. The best method is \mbox{WCR-S} in every
case. It performs remarkably well for $G \geq 30$. However, all three
WCR methods perform well for the larger values of $G$. Indeed, by most
standards, every method shown in Panel~(b) of \Cref{fig:4} works very
well, unless $G$ is less than about~30. For $G=84$, CV$_{\tn3}$ is the
worst method, but even it rejects only 5.49\% of the time. For
comparison, CV$_{\tn1}$ rejects 9.04\% of the time, and CV$_{\tn2}$
rejects~7.15\%. The best method, \mbox{WCR-S}, rejects 4.97\% of the
time.

Many applications of cluster-robust inference involve treatment at the
cluster level, and existing methods generally perform very poorly when
either the number of treated clusters or the number of control
clusters is small. Using CV$_{\tn1}$ with the $t(G-1)$ distribution or
\mbox{WCU-C} leads to severe over-rejection, and using \mbox{WCR-C} leads
to severe under-rejection \citep{MW-JAE, MW-EJ}. Our next set of
experiments therefore focuses on the model
\begin{equation}
\label{eq:treat}
y_{gi} = \beta_1 + \biZ_{gi}\bbeta_2 + \beta_k x_g + u_{gi},
\end{equation}
where $x_g$ is a treatment dummy, $\biZ_{gi}$ is a row vector of other
regressors, and $u_{gi}$ is generated by a random-effects model with
intra-cluster correlation~$\rho$. The treatment dummy equals~1 for
$G_1$ of the $G$ clusters and~0 for the remaining $G_0=G-G_1$. The
clusters that are treated are chosen at random. The $\biZ_{gi}$
consist of eight more dummy variables. For each of these variables and
each cluster, a probability $\pi_g$ between 0.25 and 0.75 is chosen at
random for each replication. Then each observation for that variable
in that cluster equals~1 with probability $\pi_g$ and~0 otherwise.
Thus all the regressors are dummies, which vary at the individual
level in a way that varies across clusters.

\begin{figure}[tb]
\begin{center}
\caption{Rejection frequencies based on $t(G-1)$ distribution 
for treatment regression}
\vspace*{-0.5em}
\label{fig:5}
\includegraphics[width=0.95\textwidth]{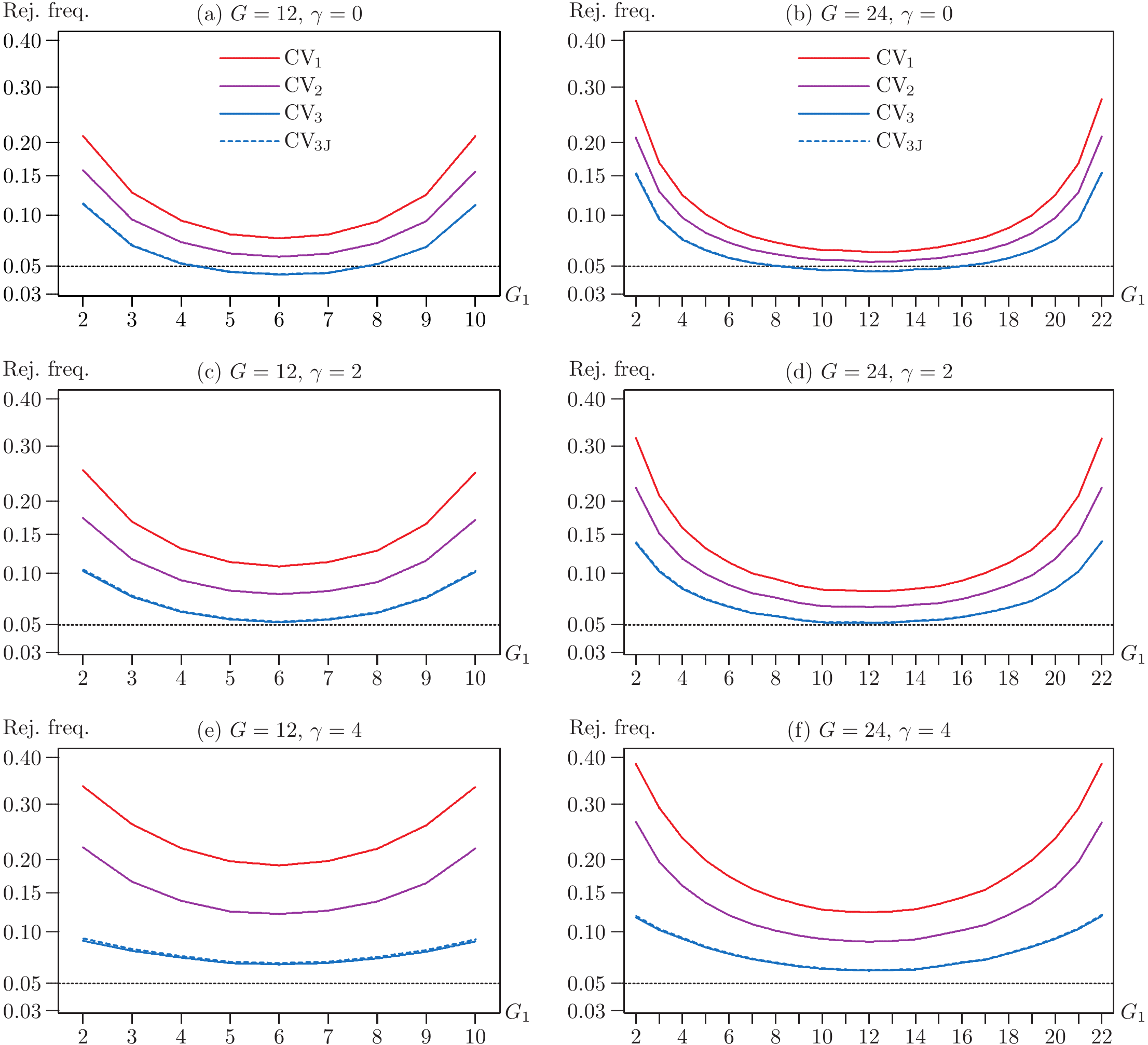}
\end{center}
{\footnotesize \textbf{Notes:} The vertical axes, which have been
subjected to a square\tkk-root transformation, show rejection frequencies
for tests of $\beta_k=0$ in \eqref{eq:treat} at the .05 level. The
horizontal axes show $G_1$, the number of treated clusters. Results
are based on $400,\tn000$ replications, with $k=10$ regressors and
$\rho=0.10$. There are either 12 or 24 clusters, with 400 observations
per cluster on average. Treated clusters are chosen at random.}
\end{figure}

\Cref{fig:5} shows rejection frequencies based on the $t(G-1)$
distribution. In the left-hand column, there are 12 clusters and 4800
observations. In the right-hand one, there are 24 clusters and 9600
observations. The value of $\gamma$ is 0 in the top row, 2 in the
middle row, and 4 in the bottom row. The number of treated
observations $G_1$ varies between 2 and $G-2$ on the horizontal axes.
It would have been impossible to set $G_1=1$ or $G_1=G-1$, because
CV$_{\tn2}$, CV$_{\tn3}$, and CV$_{\tn3{\rm J}}$ cannot be computed in
those cases. This is obvious from \eqref{eq:delone} for the
jackknife\tkk-based estimators. When the single treated cluster is
omitted, the coefficient of interest in $\hat\bbeta^{(g)}$ is not
identified.

As previous work has shown, tests that use CV$_{\tn1}$ tend to
over-reject severely when either $G_0$ or $G_1$ is small. This is
evident in \Cref{fig:5}. The over-rejection is worst in Panel~(f),
where both $\gamma$ and $G$ are largest. CV$_{\tn2}$ over-rejects less
than CV$_{\tn1}$, but it still does not work very well, except perhaps
for values of $G_1$ near $G/2$ when $\gamma=0$; see Panels (a) and~(b).
In contrast, CV$_{\tn3}$ and CV$_{\tn3{\rm J}}$, which perform
almost identically, are much less prone to over-reject than the other
two CRVEs. They actually under-reject for values of $G_1$ fairly near
$G/2$ when $\gamma=0$, and they perform very well for values of $G_1$
near $G/2$ when $\gamma=2$. Oddly, CV$_{\tn3}$ and CV$_{\tn3{\rm J}}$
over-reject less seriously for extreme values of $G_1$ when $\gamma$
is large than when $\gamma$ is small.

\begin{figure}[tb]
\begin{center}
\caption{Bootstrap rejection frequencies for treatment regression}
\vspace*{-0.5em}
\label{fig:6}
\includegraphics[width=0.95\textwidth]{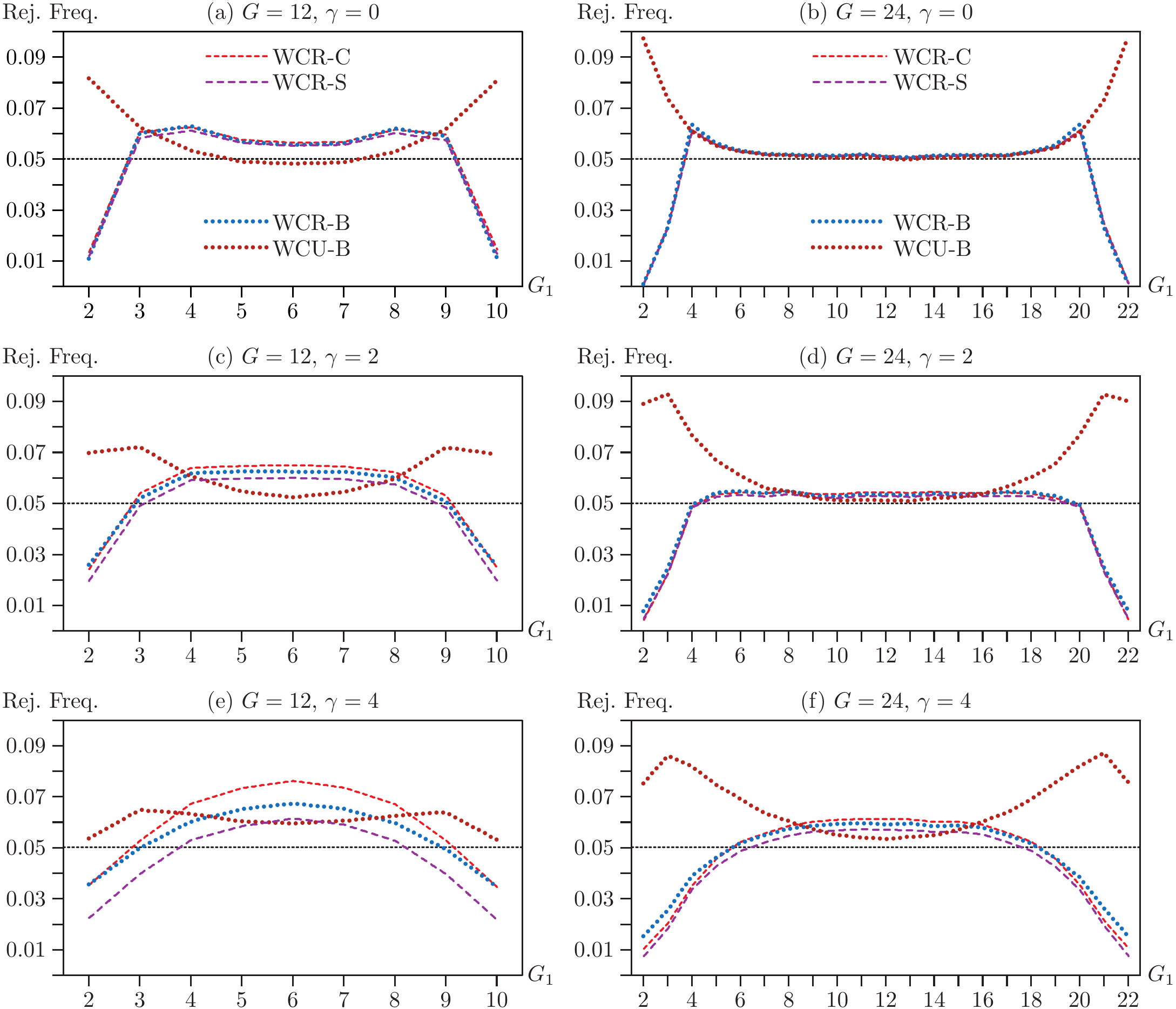}
\end{center}
{\footnotesize \textbf{Notes:} The vertical axes show rejection 
frequencies for tests of $\beta_k=0$ in \eqref{eq:treat} at the .05
level. The horizontal axes show $G_1$, the number of treated clusters.
Results are based on $400,\tn000$ replications, with $k=10$, $\rho=0.10$,
and $B=399$ bootstrap samples. There are either 12 or 24 clusters,
with 400 observations per cluster on average.}
\end{figure}

\Cref{fig:6} shows results for four bootstrap tests for the same set
of experiments as in \Cref{fig:5}. When $\gamma=0$, all three variants
of the WCR bootstrap perform almost identically. However, as $\gamma$ 
increases, their performance starts to differ. \mbox{WCR-S} seems to 
reject least frequently, which is a good thing for intermediate values
of $G_1$ and a bad thing for extreme values. In contrast, \mbox{WCR-B} 
under-rejects least severely for extreme values of~$G_1$. However, for
intermediate values, it over-rejects less than \mbox{WCR-C} but more
than \mbox{WCR-S}.

The most surprising results in \Cref{fig:6} are the ones for the
unrestricted wild bootstraps. We do not report results for \mbox{WCU-C}
or \mbox{WCU-S}, because they would have required a much longer vertical
axis. \mbox{WCU-C} rejects almost 28\% of the time in its worst case
($G=24$, $G_1=2$, $\gamma=4$), and \mbox{WCU-S} rejects over 12\% of
the time in its worst case ($G=24$, $G_1=2$, $\gamma=0$). In contrast,
\mbox{WCU-B} is arguably the best method overall when $G=12$, and it
performs very well for intermediate values of $G_1$ when $G=24$. In
addition, it never over-rejects as severely as CV$_{\tn3}$ for extreme
values of~$G_1$.

Simulations in \citet{DMN_2019} suggest that many methods work poorly
when one cluster is much bigger than the others. Even when $\gamma=4$, 
the largest cluster in our experiments is never dramatically larger 
than all the rest, although this happens quite often in empirical work. 
For instance, more than half of all incorporations in the United States 
occur in Delaware \citep{HS_2020}. This implies that studies of the 
effects of corporate governance based on changes in state laws, where 
standard errors are clustered by state of incorporation, are likely to 
encounter severe errors of inference. To investigate this phenomenon, 
we create artificial samples with 50 clusters based on data for 
incorporations by year and state from \citet{SW-data}. There are 
$205,\tn566$ observations, of which $108,\tn538$, or 52.80\%, are for 
Delaware. The second-largest cluster is Nevada, with $17,\tn010$ or 
8.27\%, and the smallest is Montana, with 101 or~0.05\%.

We perform a set of experiments similar to the ones in
\Cref{fig:5,fig:6} using these artificial samples. There are 10
regressors, generated in the same way as before, with one exception.
Because investigators are surely aware of whether or not the largest
cluster (Delaware) is treated, it is always treated in our
experiments. The other clusters to be treated (between 1 and 47 of
them) are chosen at random. Because the largest cluster is always
treated, the rejection frequencies are no longer the same for $G_1$
and $G-G_1$ treated clusters. However, since this is a pure treatment
model, the results for $G_1$ treated clusters that include Delaware
must be the same as the results for $G-G_1$ treated clusters that
exclude Delaware.

\begin{figure}[tb]
\begin{center}
\caption{Rejection frequencies when a treated cluster is very large}
\vspace*{-0.5em}
\label{fig:7}
\includegraphics[width=0.90\textwidth]{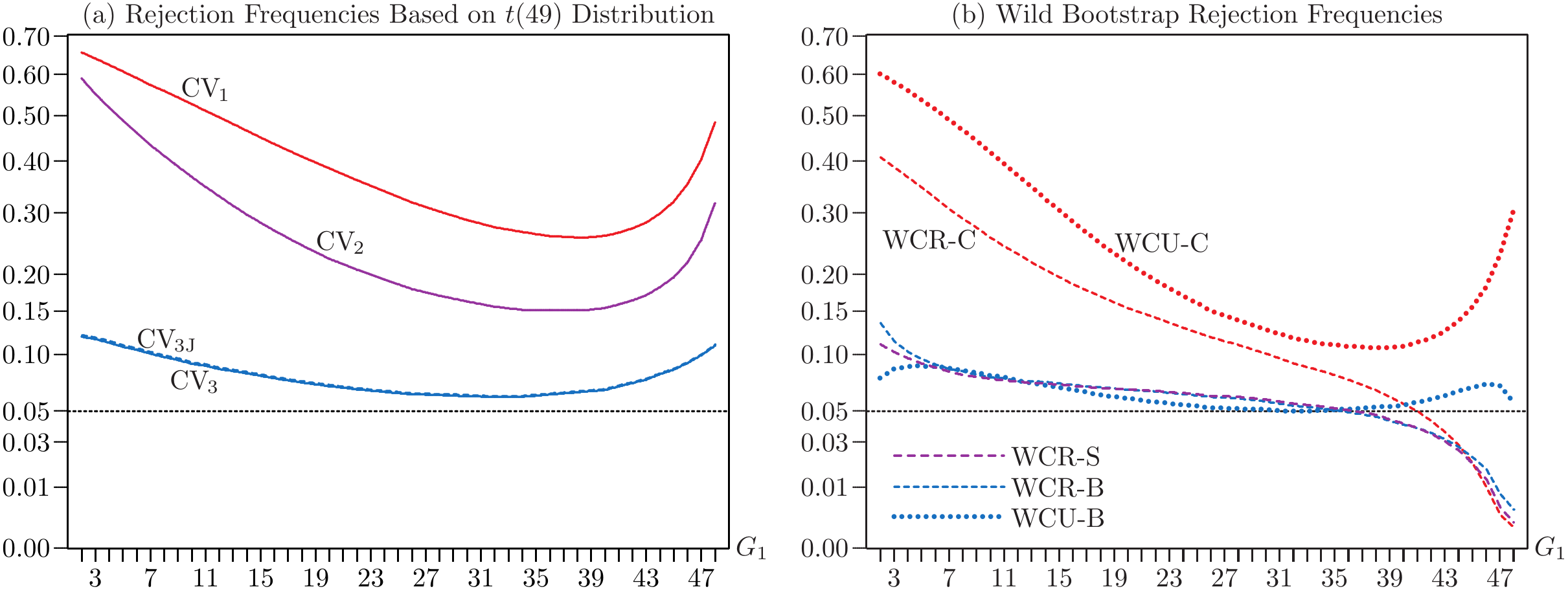}
\end{center}
{\footnotesize \textbf{Notes:} The vertical axes show rejection 
frequencies for tests of $\beta_k=0$ in \eqref{eq:treat} at the .05
level. Results are based on $400,\tn000$ replications, with $k=10$,
$\rho=0.10$, and $B=399$. There are $205,\tn566$ observations and 50
clusters, with cluster sizes proportional to incorporations in U.S.
states. The largest cluster is always treated, and the other clusters
are treated at random. The number of treated clusters varies from 2 to
14 by 1, from 16 to 36 by 2, and then from 38 to 48 by~1.}
\end{figure}

The results in \Cref{fig:7} are striking. In Panel~(a), using either
CV$_1$ or CV$_2$ leads to over-rejection that varies between severe
and extreme. Using CV$_3$ and CV$_{3{\rm J}}$ also leads to
over-rejection, but it is much less severe. For between 20 and 41
treated clusters, rejection frequencies are less than~0.07. In 
Panel~(b), \mbox{WCU-C} over-rejects severely, and \mbox{WCR-C} can either
over-reject or under-reject, often severely. In contrast, our new
bootstrap methods work remarkably well. The best of them is
\mbox{WCU-B}, which always rejects less than 9\% of the time and
sometimes rejects just about 5\% of the time. \mbox{WCR-S} and
\mbox{WCR-B} also perform much better than \mbox{WCR-C}, except when $G_1$
is very large, in which case they under-reject severely.

Even though it is based on real data, the distribution of cluster
sizes in the experiments of \Cref{fig:7} is very extreme. The
performance of CV$_3$ and three of our new bootstrap methods is far
from perfect, but it is generally very much better than that of
existing methods. Thus jackknife\tkk-based methods seem to be
remarkably robust to heterogeneity in cluster sizes.

\subsection{Test Power}
\label{subsec:power}

It is natural to worry that a new test may be less powerful than
existing tests, especially when it performs much better under the null
hypothesis. In this section, we therefore investigate test power. 
Studying power is tricky, because it is unreasonable to compare tests 
that have noticeably different rejection frequencies under the null.
If, for example, an asymptotic test rejects 15\% of the time under the
null and a bootstrap test rejects 6\% of the time, then we would
expect the asymptotic test to have substantially more power than the
bootstrap test. But the additional power may be entirely spurious,
simply reflecting the finite\tkk-sample over-rejection by the former.

One way to compare tests with different rejection frequencies under 
the null is to ``size\tkk-adjust'' them. But this approach has two
serious conceptual difficulties. First, size\tkk-adjusted tests are
infeasible. What do we learn by comparing tests that cannot actually
be performed? Second, there are often many ways to size\tkk-adjust a
given test, and they may yield quite different results. The idea of
size\tkk-adjustment is to base rejection frequencies for tests under
the alternative on critical values calculated by simulation under the
null. But, in general, there exists an infinite number of DGPs that
satisfy the null hypothesis. If they all yield the same critical
values, then there is no problem. But if they yield different critical
values, as will often be the case, then we have to choose which null
DGP to use. It seems natural to make the null DGP used for critical
values as close as possible to the alternative DGP. \cite{DM_2006}
suggests a particular way of doing this, based on the Kullback-Leibler
information criterion, but this approach means using a different
critical value for each set of values of the parameters under test.

\begin{figure}[tb]
\begin{center}
\caption{Power functions for several tests}
\vspace*{-0.5em}
\label{fig:8}
\includegraphics[width=0.90\textwidth]{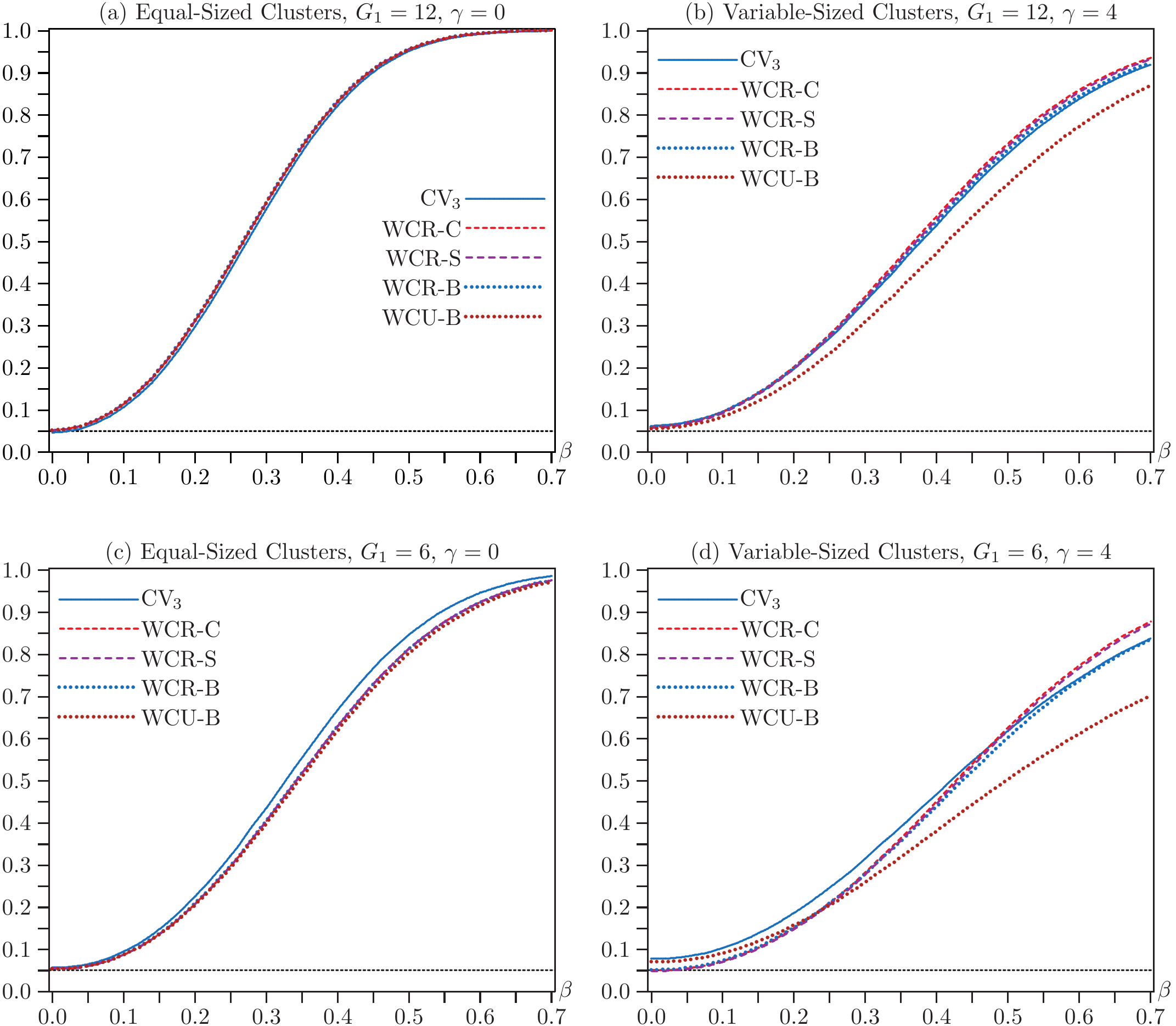}
\end{center}
{\footnotesize \textbf{Notes:} The vertical axes show rejection 
frequencies for tests at the .05 level. Results are based on $400,\tn000$
replications, with $G=24$, $N=9600$, $k=5$, $\rho=0.10$, and $B=999$.
The hypothesis being tested is $\beta_k=0$ in \eqref{eq:treat}. The
horizontal axes show the values of $\beta$ in the DGP.}
\end{figure}

To avoid the difficulties just discussed, we focus on four cases where
the tests of interest all perform quite well under the null. They are 
treatment experiments similar to the ones in \Cref{fig:5,fig:6}, with 
$G=24$, $N=9600$, and $k=5$. In Panels (a) and~(b), $G_1=12$, so that
precisely half the clusters are treated. In Panels (c) and~(d),
$G_1=6$, so that the effects of having few treated clusters are
apparent but not severe. In order to avoid excessive power loss, we
use $B=999$ for the bootstrap tests. We use $k=5$ instead of $k=10$
partly to reduce computational cost and partly to improve test
performance under the null.

\Cref{fig:8} shows rejection frequencies as a function of $\beta_k$,
the actual coefficient on the treatment dummy in \eqref{eq:treat}, when
the null hypothesis is that $\beta_k=0$. In Panels~(a) and~(c),
$\gamma=0$, so that every cluster has exactly 400 observations. In
Panel~(a), the perfectly balanced case, all five power functions are
visually indistinguishable. In Panel~(c), where only six clusters are
treated, CV$_3$ has noticeably more power than any of the bootstrap
methods, which are all but identical.

In Panels (b) and~(d), cluster sizes vary from 32 to~1513. All tests
are now substantially less powerful than in Panels (a) and~(c),
because, whenever there is intra-cluster correlation, the information
content of a sample declines as the cluster sizes become more
variable. The most striking result in both panels is that \mbox{WCU-B}
has noticeably less power than any of the other methods. This is
especially true in Panel~(d), where \mbox{WCU-B} over-rejects modestly
under the null but becomes by far the least powerful method for larger
values of~$\beta_k$.

The pattern for CV$_{\tn3}$ is similar but much less pronounced. Under
the null hypothesis, it over-rejects slightly under the null in
Panel~(b) and noticeably in Panel~(d), with rejection frequencies of
0.0612 and 0.0775, respectively. But for large enough values
of~$\beta_k$, it has less power than \mbox{WCR-C} and \mbox{WCR-S},
especially in Panel~(d). The latter two methods also have slightly
more power than \mbox{WCR-B} in Panel~(b) and noticeably more in
Panel~(d) for large values of~$\beta_k$. Interestingly, \mbox{WCR-V},
which for clarity is not shown in the figure, has somewhat less power
than either \mbox{WCR-C} or \mbox{WCR-S} in Panels~(b) and~(d), where
cluster sizes vary a lot. In contrast, it is almost indistinguishable
from both these methods in Panels~(a) and~(c), where cluster sizes are
constant.

Based on these results, the procedure of choice appears to be 
\mbox{WCR-S}. For larger values of~$\beta_k$, it is always one of the 
two most powerful tests. \mbox{WCR-C} has similar power, and it also 
works well under the null in these experiments, but it is much more 
prone to over-reject than \mbox{WCR-S} in 
\Cref{fig:3,fig:4,fig:6,fig:7}. Happily, \mbox{WCR-S} is already 
available in computationally efficient packages for \texttt{Stata}, 
\texttt{R}, and \texttt{Python}; see \Cref{sec:boot}.

\subsection{Confidence Intervals}
\label{subsec:CI}

Cluster-robust standard errors and bootstrap methods are often used to
form confidence intervals. Although we do not perform any Monte Carlo
experiments explicitly to study the properties of confidence
intervals, these can be inferred from \Cref{fig:8} and the results in
\Cref{subsec:size}. Most confidence intervals are implicitly or
explicitly obtained by inverting a hypothesis test. When such a test
has approximately the correct rejection frequency, the resulting
confidence interval must have approximately correct coverage.
Similarly, when such a test has high power, the resulting confidence
interval must be relatively short.

In many of the experiments in \Cref{subsec:size}, tests based on
CV$_{\tn3}$ and the $t(G-1)$ distribution are much less prone to
over-reject than tests based on CV$_{\tn1}$. This suggests that the
coverage of confidence intervals based on CV$_{\tn3}$ standard errors
will often be much better than the coverage of ones based on
CV$_{\tn1}$ standard errors. Even more reliable intervals may often
(but not always) be obtained by using the \mbox{WCR-S} or \mbox{WCR-B}
bootstraps, which perform much better than the classic \mbox{WCR-C}
bootstrap in many cases. The \mbox{WCU-B} bootstrap also performs well
in many cases under the null, but the results in Panels (b) and (d) of
\Cref{fig:8} suggest that, when cluster sizes vary a lot, intervals
based on it may be longer than ones based on \mbox{WCR-B}, which in turn
may be slightly longer than ones based on \mbox{WCR-S}.

The \mbox{WCR-S} bootstrap has excellent performance in many of the
experiments of \Cref{subsec:size}, seems to have slightly better power
than \mbox{WCR-B} in Panels~(b) and~(d) of \Cref{fig:8}, and is easy
to compute. Therefore, we tentatively recommend that confidence
intervals should be obtained by inverting \mbox{WCR-S} bootstrap
tests. However, using CV$_{\tn3}$ standard errors and the $t(G-1)$
distribution would often lead to very similar intervals. 

Of course, it is easier to obtain a confidence interval by using a
standard error and the $t(G-1)$ distribution than by inverting a
bootstrap test, and it is easier to invert any form of WCU bootstrap
test than any form of WCR bootstrap test. However, the computational
cost of inverting WCR bootstrap tests can be remarkably small, even
for very large samples; see \citet[Section~3.5]{RMNW} and 
\citet[Section~3.4]{JGM-fast}.

\section{Empirical Examples}
\label{sec:empirical}

In this section, we consider three empirical examples. These suggest
that the new bootstrap procedures proposed in \Cref{sec:boot} may
sometimes yield results very similar to those from the existing
\mbox{WCR-C} and \mbox{WCU-C} procedures, but they may also yield 
results which differ noticeably from those and from each other.

\subsection{Minimum Wages and Hours Worked}
\label{subsec:minwage}

Our first example is based on \citet[Section~8]{MNW-guide}. It
exploits differences in the minimum wage across states and years to
estimate the impact of minimum wages on hours worked for teenagers.

The data on hours at the individual level from the American Community
Survey (ACS) are obtained from IPUMS \citep{IPUMS_2020} and cover the
years 2005\tkk--2019. The minimum wage data come from
\cite{Neumark_2019} and are collapsed to state\tkk-year averages to
match the ACS frequency. We restrict attention to teenagers aged
16\tkk--19, keeping only individuals who are children of the
respondent to the survey and who have never been married. We drop
individuals who had completed one year of college by age~16 and those
reporting in excess of 60~hours usually worked per week. We also
restrict attention to individuals who identify as either black or
white. There are $492,\tn827$ observations in 51 clusters, which
correspond to all 50 states plus the District of Columbia.

The model we estimate is
\begin{equation}
\label{reg:wage}
y_{ist} = \alpha + \beta\tk \textrm{mw}_{\tn st} + \biZ_{ist}\bgamma
+ \delta_s + \eta_t + u_{ist},
\end{equation}
where $y_{ist}$ is usual hours worked per week for individual~$i$. The
parameter of interest is $\beta$, which is the coefficient on
$\textrm{mw}_{\tn st}$, the minimum wage in state $s$ at time~$t$. The
row vector $\biZ_{ist}$ collects a large set of individual-level
controls, including race, gender, age, and education. There are also
state and year fixed effects, denoted by $\delta_s$ and $\eta_t$,
respectively.

As \citet{MNW-guide} discusses, clustering could in principle be done
at several different levels. However, the one that is most appealing
and seems to be supported by the data is clustering at the state
level. The 51 clusters vary considerably in size. The smallest has 258
observations, and the largest has $35,\tn995$. The ratio of these
numbers is more than twice as large as for $\gamma=4$ in the
experiments of \Cref{subsec:size}. The mean number of observations per
cluster is $9,\tn663$, and the median is $7,\tn082$. This suggests
that inference based on CV$_{\tn1}$ and the $t(50)$ distribution may
not be reliable. Other measures of cluster heterogeneity, which are
discussed in the original paper, lead to the same conclusion.

\begin{table}[tp]
\caption{Example 1, minimum wages and hours worked}
\label{tab:hours}
\vspace*{-2em}
\begin{center}
\begin{tabular*}{0.85\textwidth}{@{\extracolsep{\fill}}lrcrc}
\toprule
& Estimate & Std.\ error & $t$-statistic & $P$ value \\
\midrule
HC$_1$      & $-0.15389$ & $0.02825$ & $-5.4471$ & $0.0000$ \\
CV$_{\tn1}$ & $-0.15389$ & $0.06231$ & $-2.4697$ & $0.0170$ \\
CV$_{\tn3}$ & $-0.15389$ & $0.06713$ & $-2.2925$ & $0.0261$ \\
\midrule
\multicolumn{5}{l}{Wild cluster bootstrap $P$ values} \\
\midrule
WCR-C & $0.0362$ & WCU-C & $0.0207$\\
WCR-V & $0.0352$ & WCU-V & $0.0186$\\
WCR-S & $0.0374$ & WCU-S & $0.0227$\\
WCR-B & $0.0371$ & WCU-B & $0.0203$\\
\bottomrule
\end{tabular*}
\vskip 4pt
\parbox{0.85\textwidth}{\footnotesize
\textbf{Notes:} There are $492,\tn827$ observations, 51
clusters, and 79 coefficients, including state and year fixed effects.
The coefficient of interest is $\beta$ in~\eqref{reg:wage}.
Bootstrap $P$ values use $B=999,\tn999$.}
\end{center}
\end{table}

\Cref{tab:hours} presents our key results. As expected, the
CV$_{\tn3}$ $t$-statistic is somewhat smaller than the CV$_{\tn1}$
$t$-statistic, and the $P$~value based on the $t(50)$ distribution is
therefore somewhat larger. The four WCR $P$~values are larger than
either of them, but still below~0.05, and the four WCU $P$~values are
notably smaller than the WCR ones. Because $B$ is so large (larger
than really needed), the simulation standard errors for the
WCR bootstrap $P$~values are about~0.0002.

Based on how similar the four WCR $P$~values are, and on how well many
of the WCR methods perform in the experiments of \Cref{subsec:size},
we tentatively conclude that the ``true'' $P$~value for the test of
$\beta=0$ is probably between 0.034 and~0.039. Thus the null
hypothesis can safely be rejected at the 0.05 level but not at the
0.01 level.

\subsection{Political Turnover and Test Scores}
\label{subsec:turnover}

The second example comes from \citet*{AMT_2022}. This paper examines
the impact of political turnover on the quality of public services.
Specifically, it examines several outcomes following close mayoral
elections in Brazil. One of these outcomes is the test scores of
fourth-grade students. The paper uses a regression discontinuity
design to identify the treated and control municipalities, but it
conducts the analysis using OLS.  We replicate one such regression,
found in Table~3, Column~5 of the original paper:
\begin{equation}
   \textrm{score}_{imt+1} = \alpha + \beta\tk\IF(\textrm{IVM}_{mt}<0)
   + \gamma\tk\textrm{IVM}_{mt}
   + \delta\tk\IF(\textrm{IVM}_{mt}<0) \textrm{IVM}_{mt} 
   + \eta\tk\textrm{score}_{imt} +  \epsilon_{imt}.
\label{reg:turnover}
\end{equation}
The dependent variable is the test score one year after an election.
$\textrm{IVM}_{mt}$ is the incumbent vote margin in the close election
which occurs in year $t$. Accordingly, the treatment variable is
$\IF(\textrm{IVM}_{mt}<0)$, which equals~1 when the incumbent party
loses the election and a turnover occurs, and the coefficient of
interest is~$\beta$. This regression is estimated using a sample which
is determined by a selected bandwidth. While the paper considers
several bandwidths, we focus on the bandwidth 0.110, as this results in 
the largest sample.

The paper clusters the standard errors at the municipality level.
Since there are 2101 municipalities, many of them located close
to each other, it seems possible that this level of clustering is too
fine. We therefore consider state-level clustering. However, there are
only 26 states in Brazil, and they vary in size from 420 to
$64,\tn953$ with partial leverages from 0.000234 to 0.179318
\citep{MNW-influence}. With this much heterogeneity across clusters,
relying on CV$_{\tn1}$ may be risky.

\begin{table}[tp]
\caption{Example 2, political turnover and test scores}
\label{tab:turnover}
\vspace*{-2em}
\begin{center}
\begin{tabular*}{0.85\textwidth}{@{\extracolsep{\fill}}lrcrc}
\toprule
& Estimate & Std.\ error & $t$-statistic & $P$ value \\
\midrule
HC$_1$      & $-0.06684$ & $0.00528$ & $-12.6616$ & $0.0000$ \\
CV$_{\tn1}$ (munic.) & $-0.06684$ & $0.02430$ & $-2.7505$ & $0.0060$ \\
CV$_{\tn1}$ & $-0.06684$ & $0.02204$ & $-3.0326$ & $0.0056$ \\
CV$_{\tn3}$ & $-0.06684$ & $0.02411$ & $-2.7722$ & $0.0104$ \\
\midrule
\multicolumn{5}{l}{Wild cluster bootstrap $P$ values} \\
\midrule
WCR-C & $0.0047$  & WCU-C & $0.0193$\\
WCR-V & $0.0057$  & WCU-V & $0.0235$\\
WCR-S & $0.0046$  & WCU-S & $0.0212$\\
WCR-B & $0.0056$  & WCU-B & $0.0236$\\
\bottomrule
\end{tabular*}
\vskip 4pt
\parbox{0.85\textwidth}
{\footnotesize \textbf{Notes:} There are $429,\tn979$ observations, 26
clusters, and 5 coefficients. The coefficient of interest is $\beta$ 
in~\eqref{reg:turnover}. Bootstrap $P$ values use $B=999,\tn999$.}
\end{center}
\end{table}

\Cref{tab:turnover} presents our key results. As expected, the 
CV$_{\tn1}$ standard error for clustering by state is smaller than the
CV$_{\tn3}$ standard error. Contrary to our expectations, however,
both are a bit smaller than the CV$_{\tn1}$ standard error for
clustering by municipality. The four WCR $P$ values are similar to
each other and to the $P$ value based on the CV$_{\tn1}$ $t$-statistic
and the $t(25)$ distribution. Surprisingly, the four WCU $P$ values
are noticeably larger than the WCR ones. Nevertheless, since every
test rejects at the 0.05 level, there is evidence against the null 
hypothesis.

\subsection{Patronage in the British Empire}
\label{subsec:patronage}

The third example is taken from \cite{Xu_2018}, which explores the
effect of patronage in the colonial era of Britain on the appointment
of governors to colonies.  Part of the analysis examines whether the
extent to which the current secretary of state and a governor are
``connected'' led to more desirable colony postings. We replicate the
results of one such regression, found in Table~3, Column~3 of the
original paper:
\begin{equation}
  \textrm{log(revenue)}_{ist} = \alpha + \beta_1
  \textrm{connected}_{it} + \beta_2 \textrm{served}_{it}  
      + \gamma_i + \tau_t + \delta_{it} + \epsilon_{ist}.
  \label{reg:patronage}
\end{equation}
Here $\textrm{log(revenue)}_{ist} $ is the initial revenue for colony
$s$ when governor $i$ was appointed in year~$t$. The main variable of
interest is $\textrm{connected}_{it}$, which is a binary variable set
equal to~1 when the governor and the secretary share connections such
as having attended the same elite boarding school, or Oxford or
Cambridge, or both being in the aristocracy, or having shared
ancestry. The variable $\textrm{served}_{it}$ is the number
colonies in which the governor has served up to the year of
appointment. The regression also has fixed effects for governors
($\gamma_i$), years ($\tau_t$), and the duration of the governorship
($\delta_{it}$).

The paper clusters the standard errors at the bilateral pair (or dyad)
level between the secretary of state and the governor. However, the 
dependent variable is observed at multiple times for each colony, so
it seems likely that there would be dependence across observations for
the same colony. The regression does not include colony fixed effects,
which would have reduced this dependence, because, with so many other
fixed effects, it was impossible to include them. Thus, it seems
plausible that the standard errors should be clustered at the colony
level instead of the dyad level, and we investigate this approach.
Switching from dyadic clustering to clustering by colony actually
reduces the CV$_{\tn1}$ standard error. However, even though there are
70 colonies, they are quite unbalanced; the number of observations per
colony ranges from 4 to~104. Partial leverages also vary greatly, and
they seem to be roughly proportional to cluster sizes. Perhaps in
consequence, the CV$_{\tn3}$ standard error is 47\% larger than the
CV$_{\tn1}$ one.

\begin{table}[tp]
\caption{Example 3, patronage in the British empire}
\label{tab:patronage}
\vspace*{-2em}
\begin{center}
\begin{tabular*}{0.85\textwidth}{@{\extracolsep{\fill}}lrcrc}
\toprule
& Estimate & Std.\ error & $t$-statistic & $P$ value \\
\midrule
HC$_1$      & $0.17722$ & $0.07573$ & $2.3401$ & $0.0193$ \\
CV$_{\tn1}$ (dyadic) & $0.17722$ & $0.09933$ & $1.7842$ & $0.0750$ \\
CV$_{\tn1}$ & $0.17722$ & $0.08702$ & $2.0366$ & $0.0455$ \\
CV$_{\tn3}$ & $0.17722$ & $0.12810$ & $1.3834$ & $0.1710$ \\
\midrule
\multicolumn{5}{l}{Wild cluster bootstrap $P$ values} \\
\midrule
WCR-C & $0.0535$ & WCU-C & $0.0575$\\
WCR-V & $0.0704$ & WCU-V & $0.0738$\\
WCR-S & $0.0656$ & WCU-S & $0.0725$\\
WCR-B & $0.0621$ & WCU-B & $0.0678$\\
\bottomrule
\end{tabular*}
\vskip 4pt
\parbox{0.85\textwidth}
{\footnotesize \textbf{Notes:} There are $3510$ observations, 70
clusters, and 573 coefficients. The coefficient of interest is $\beta_1$
in~\eqref{reg:patronage}. Bootstrap $P$~values use $B=99,\tn999$
because, with 573 regressors, the computations for WCR/WCU-V and
WCR/WCU-B are much more expensive than for the previous examples.}
\end{center}
\end{table}

In view of the dramatic difference between the CV$_{\tn1}$ and
CV$_{\tn3}$ $t$-statistics, the various wild bootstrap methods provide
valuable information. The bootstrap $P$~values are all somewhat larger
than the one for the CV$_{\tn1}$ $t$-statistic based on the $t(69)$
distribution, but they are all much smaller than the corresponding one
for the CV$_{\tn3}$ $t$-statistic. The smallest bootstrap $P$~value is
the one for the classic \mbox{WCR-C} method. At $0.0535$, it is not
much larger than the one based on the $t(69)$ distribution.
Surprisingly, every WCU $P$~value is larger than the corresponding WCR
$P$~value.

This example is deliberately extreme, because the number of regressors
(573) is unusually large relative to the number of
observations~(3510). Perhaps in consequence, the full coefficient
vectors $\hat\bbeta^{(g)}$ are not identified for 61 out of the 70
clusters. However, since the $\hat\beta_1^{(g)}$ coefficients are
always identified, we used a generalized inverse to compute both
CV$_{\tn3}$ and the bootstrap DGPs for the \mbox{WCR/WCU-S} and
\mbox{WCR/WCU-B} bootstraps. The alternative approach of trying to
estimate a variance matrix based on only 9 out of 70 clusters seems
very dubious, and it yields an implausibly small standard error of
just~0.0379. However, the large number of singularities may explain
why the CV$_{\tn3}$ and CV$_{\tn1}$ standard errors differ as much as
they do.

Because $k$ is so large in this example, we suspect that the $t$-test 
based on CV$_{\tn3}$ may be prone to under-reject, and that both the 
$t$-test based on CV$_{\tn1}$ and the \mbox{WCR-C} bootstrap test may 
be prone to over-reject; see \Cref{fig:3}. Nevertheless, the $P$~values
for the new WCR bootstrap methods are only modestly larger than the
\mbox{WCR-C} $P$~value. The fact that all the bootstrap $P$~values lie
between 0.0535 and 0.0738 suggests that the ``true'' $P$~value
probably also lies within, or at least not too far outside, this
interval. We conclude that there seems to be only weak evidence
against the null hypothesis.

\section{Conclusion and Recommendations}
\label{sec:conc}

The classic CV$_{\tn1}$ estimator given in \eqref{eq:CV1} is by far
the most popular CRVE for linear regression models, but standard
errors based on it are often much too small. The cluster jackknife
estimator, often called CV$_{\tn3}$, has been known for many years but
is much less widely used. In \Cref{sec:jack}, we discuss how to
compute CV$_{\tn3}$ in a computationally efficient fashion. Except
when all clusters are tiny, this is the fastest available method for
computing it; see \Cref{sec:speed}. Inference based on CV$_{\tn3}$ and
the Student's $t(G-1)$ distribution seems to be much more reliable
than inference based on CV$_{\tn1}$ and that distribution; see
\Cref{subsec:size}. This accords with theoretical results in
\citet{Hansen-jack}, which provides no simulations and cites the ones
in this paper.

Although combining CV$_{\tn3}$ standard errors and the $t$
distribution often works well, it does not always do so. Bootstrap
methods may well perform better, and they also provide a valuable
robustness check. In \Cref{sec:boot}, we prove some simple, but by no
means obvious, algebraic results about the relationship between
cluster jackknife estimates and score vectors at the cluster level.
These results allow us to obtain new and easy-to\tkk-compute variants
of the wild cluster bootstrap. These typically perform better than the
classic variants, now called \mbox{WCR-C} and \mbox{WCU-C}. The eight
new and existing variants are summarized in \Cref{tab:eight}. Of
these, the ones that use CV$_{\tn1}$ together with modified bootstrap
score vectors, called \mbox{WCR-S} and \mbox{WCU-S}, are particularly
easy to compute. They are available in packages for \texttt{Stata}
and~\texttt{R}.

Prior to this paper, there were already quite a few methods for
inference in linear regression models with clustered disturbances
\citep{MNW-guide}, and \Cref{sec:boot} has added six new variants of
the wild cluster bootstrap. Empiricists may reasonably ask what
methods they should use in practice. As discussed in detail in 
\cite{MNW-guide,MNW-influence}, the first thing to do is to
investigate the clustering structure of the model and dataset. For
instance, it is good practice to calculate the effective number of
clusters \citep{CSS_2017} as well as various measures of leverage and
influence at the cluster level \citep{MNW-influence}. When these
measures indicate that clusters are well-balanced, and the (effective)
number of clusters is large (say, more than~100), then CV$_{\tn1}$ and
CV$_{\tn3}$ should yield very similar standard errors. In such cases,
it is probably safe to rely on CV$_{\tn3}$ standard errors together
with the $t(G-1)$ distribution.

However, the number of clusters will often be much less than~100.
Moreover, measures of cluster-level leverage and influence may
indicate that clusters are not well-balanced. This can happen, for 
example, when cluster sizes vary a lot, when there are few treated 
clusters, or when the distributions of key regressors vary greatly 
across clusters. In such cases, CV$_{\tn1}$ and CV$_{\tn3}$ can yield
quite different standard errors. Recall \Cref{tab:patronage}, where
the CV$_{\tn3}$ standard error is 47\% larger than the CV$_{\tn1}$
standard error, even though there are 70 clusters.
Whenever CV$_{\tn1}$ and CV$_{\tn3}$ differ
substantially, bootstrap $P$~values or confidence intervals are likely
to be more reliable than conventional ones based on either of those
CRVEs, and it is probably a good idea to compute both WCR-C and WCR-S
$P$~values.

In most cases, it is advisable to compute wild cluster bootstrap 
$P$~values and/or confidence intervals using at least $9,\tn999$
bootstrap samples. This is usually not computationally difficult.
However, there might be exceptions when either the number of clusters
or the number of regressors is unusually large. Of course, when the
number of clusters is very large, the bootstrap will not be needed
unless the clusters are severely unbalanced, but that can happen.

If we had to recommend just one method, it would be the \mbox{WCR-S}
bootstrap proposed in \Cref{sec:boot}. This method uses ordinary
CV$_{\tn1}$ standard errors, which makes it easy to compute, but the
bootstrap DGP employs restricted scores that have been transformed
using the cluster jackknife. In some of our experiments, the
\mbox{WCR-S} bootstrap works substantially better than the classic (and 
popular) \mbox{WCR-C} bootstrap; see, in particular, 
\Cref{fig:2,fig:3,fig:4} and \Cref{fig:7}. We generally do not recommend
using the unrestricted wild cluster bootstrap, except perhaps as a 
robustness check or when it is desired to generate a large number of 
confidence intervals using just one set of bootstrap samples.

\setlength{\bibsep}{0pt}
\bibliography{mnw-knife}
\addcontentsline{toc}{section}{\refname}

\end{document}